# Co-design for Trustworthy AI: An Interpretable and Explainable Tool for Type 2 Diabetes Prediction Using Genomic Polygenic Risk Scores

Ralf Beuthan (4), Megan Coffee (6), Heejin Kim (1), Na Yeon Kim (1), Pedro Kringen (9), Elisabeth Hildt (2) (13), Haekyung Lee (8), Seunggeun Lee (1), Emilie Wiinblad Mathez (5), Sira Maliphol (12), Vadim Pak (11), Yuna Park (1), Stephan Sonnenberg (7), Jesmin Jahan Tithi (10), Magnus Westerlund (3), Roberto V. Zicari (1).

(1) Graduate School of Data Science, Seoul National University (SNU), Seoul, S. Korea

(2) Center for the Study of Ethics in the Professions, Illinois Institute of Technology, Chicago, USA

(3) Laboratory for Trustworthy AI, Arcada University of Applied Sciences, Helsinki, Finland

(4) Department of Philosophy at Myongji University, Seoul, S. Korea

(5) Z-inspection® Initiative, Geneva, Switzerland

(6) Department of Medicine and Division of Infectious Diseases and Immunology, NYU Grossman School of Medicine, New York, USA

(7) Seoul National University School of Law, Seoul National University (SNU), Seoul, S. Korea

(8) Division of Nephrology, Department of Internal Medicine, Soonchunhyang University Seoul Hospital, Seoul, S. Korea

(9) Trustworthy AI Lab, Østfold University College, Fredrikstad, Norway

(10) Intel Corporation, Santa Clara, California, USA

(11) Council of Europe, Administrator in the Committee on Artificial Intelligence/Steering Committee on New and Emerging Digital Technologies, France

(12) Graduate School of Engineering Practice, Seoul National University (SNU), Seoul, S. Korea

(13) L3S Research Center, Leibniz University of Hannover, Germany

# Contents

# Abstract

The polygenic risk scores (PRS) have emerged as an important methodology for quantifying genetic predisposition to complex traits and clinical disease. Significant progress has been made in applying PRS to conditions such as obesity, cancer, and type 2 diabetes (T2DM). Studies have demonstrated that PRS can effectively identify individuals at high risk, thereby enabling early screening, personalized treatment, and targeted interventions for diseases with a genetic predisposition.

One current limitation of PRS, however, is the lack of interpretability tools. To address this problem for T2DM, researchers at the Graduate School of Data Science at the Seoul National University introduced eXplainable PRS (XPRS). This visualization tool decomposes PRSs into gene-level and single-nucleotide polymorphism (SNP) contribution scores via Shapley Additive Explanations (SHAP), providing granular insights into the specific genetic factors driving an individual's risk profile. In this use case, we used a co-design approach to assess the trustworthiness of the XPRS system by considering the legal, medical, ethical, and technical robustness aspects of the AI system at the early design stage and during its potential implementation and use by medical professionals. For that, we used Z-inspection®, an ethically aligned Trustworthy AI co-design methodology, and piloted the Council of Europe's Human Rights, Democracy, and the Rule of Law Impact Assessment for AI Systems (HUDERIA) (Council of Europe (CAI) 2025).

The findings of this use-case comprise a comprehensive set of ethical, legal, and technical lessons learned. These insights, identified by a multidisciplinary team of experts (ethics, legal, human rights, computer science, and medical), serve as a framework for designers to navigate future challenges with this and other AI systems. The findings also provide a useful reference for researchers developing explainability frameworks for PRS in diverse clinical contexts.

# I. INTRODUCTION

Type 2 diabetes mellitus (T2DM) is a growing public health challenge. The International Diabetes Federation estimates that global diabetes prevalence was 10.5% in 2021 and will rise to 12.2% by 2045 [1]. This trajectory, driven by aging populations and shifting lifestyle patterns, imposes a burden on healthcare systems, increasing morbidity and mortality directly and amplifying the risks of many other conditions.

Risk prediction for T2DM has traditionally relied on clinical and lifestyle factors, and multiple validated risk equations have been developed to estimate the future onset of T2DM in routine-care settings [2,3] More recently, machine-learning models trained on large-scale health datasets have demonstrated high predictive accuracy for incident T2DM, typically achieving high Areas Under the Curve (AUC) or Concordance-statistics (C-statistics) when predicting which individuals will develop T2DM during follow-up among those normal glucose tolerance [4]. In addition, many of these models support feature-attribution analyses, which help identify key contributing risk factors and improve model interpretability.

However, genetic susceptibility strongly influences T2DM risk; a large multi-ancestry genome-wide association study (GWAS) has identified numerous loci contributing to T2DM risk [5]. Accordingly, polygenic risk scores (PRS) can quantify an individual's genetic susceptibility, and trans-ancestry approaches have been developed to improve cross-population validity [6], though significant generalizability limitations remain. Across multiple ancestry cohorts, evidence suggests that a T2DM PRS can identify individuals at high risk and improve or complement T2DM prediction beyond standard clinical risk factors [7-10]

Being able to predict disease risk and response to treatment would improve clinical outcomes in medicine. The PRS can complement usual clinical risk factors, such as age, sex and family history, behavior, biochemical biomarkers and radiological imaging [11].

As large national DNA databases are becoming increasingly available for GWAS, development of PRS, usually based on single-nucleotide polymorphisms (SNPs), has gained popularity as a method for assessing genetic susceptibility to diseases and predicting disease risks and selecting the most effective treatment [11], [12]

As explained in the article "Clinical utility of polygenic risk scores: a critical 2023 appraisal" [14]: the heritability of diseases (such as cancer or diabetes) has not been found to be explicable by a few genetic variants with strong effects, but rather hundreds to thousands of weak to moderate disease associations. The PRSs therefore sum up a large number of such single-variant association statistics so as to combine these (individually weak) effects in a single number for use in disease diagnosis, prognosis or treatment, and in research. In the process, the GWAS may be shared without raising data privacy concerns [14].

Despite these advances, PRS is often delivered as a single score, and its clinical translation is frequently limited by the lack of tools that explain which genes and variants are driving an individual's genetic risk [15].

Despite their potential, PRS often face limited clinical use. Research has shown that it may be largely because they can be difficult to interpret and may lack the necessary predictive power [12]. As pointed out by Koch et al. [14], "The diagnostic and prognostic performance of PRSs alone is consistently low and usually only accounts for the heritable component of a trait and ignores the etiological role of environment and lifestyle".

Because explainability can increase the transparency and reliability of predictions and strengthen user trust, especially in clinical decision-making contexts and where there is a vast array of different contributing risks, PRS systems need an interpretable layer that supports transparent risk communication. To address this interpretability gap, the research team at the Graduate School of Data Science at the Seoul National University designed an explainable PRS solution called XPRS.

## A. AI Solution: Function and aims of the AI based XPRS tool

XPRS is a post-hoc interpretation and visualization tool that decomposes PRSs into genes/regions and SNP contribution scores via Shapley additive explanations (SHAPs) [16], which provide insights into specific genes and SNPs that significantly contribute to the PRS of an individual [14], [12], [17].

XPRS provides multilevel visualizations (e.g., Manhattan-style summaries and Locus Zoom-like views) to highlight important genes and variants at both the population and individual level. By bridging PRS outputs, and stakeholder-understandable explanations, XPRS is intended to improve communication of genetic risk information between clinicians and patients.

The outputs of the XPRS require three input files: a genotype file (binary PLINK format), a PRS scoring file, and a GWAS association file [14], [12], [17].

XPRS is trained and validated on large East-Asian cohorts (Korean Genome and Epidemiology Study [KoGES] + Biobank Japan) and cross-validated using external hold-out sets (United Kingdom Biobank East Asian cohort) [14], [12], [17].

While the PRS system is intended primarily for population screening, it can also enable healthcare providers to proactively identify individuals who may require closer monitoring, lifestyle interventions, or early pharmacotherapy. The XPRS aims to empower both clinicians and patients to make informed, individualized healthcare decisions, ultimately aiming to reduce the risk of T2DM and its associated complications by making the genetic factors transparent.

An example use case would be the KoGES. If a participant was found to have the highest T2DM PRS in the cohort XPRS outputs would include a density plot showing this individual at the extreme high-risk tail, a Manhattan plot highlighting genes like CDKAL1 [18], and a Locus Zoom-like visualization pinpointing specific SNPs that contribute most strongly, positively or negatively, to the PRS.

XPRS Benefit Claims [15]:

- **Enhanced PRS Interpretability:** XPRS decomposes PRS into gene/region and SNP contribution scores via SHAPs, providing clear insights into genetic factors influencing disease risk.

- **Comprehensive Visualization Tools:** The software offers multilevel visualizations, including Manhattan plots and LocusZoom-like plots, for population and individual analyses, facilitating effective communication between clinicians and patients.

- **Integration with Genomic Resources:** XPRS leverages genomic annotations such as refGene and cS2G mapping to ensure precise mapping of SNPs to their corresponding genes, increasing the reliability of PRS interpretations.

- **Increased explainability and increased communication and understanding** between patients and medical professionals.

Possible contexts where the system could be used by clinicians are:

1. **Discuss Personalized Risk:** Clearly explain why this patient is at particularly high genetic risk.

2. **Recommend Targeted Interventions:** Suggest more frequent glucose testing, an intensive lifestyle program, or earlier consideration of pharmacotherapy.

3. **Monitor More Closely:** Plan regular follow-up visits to track blood sugar levels and adjust interventions as needed.[1]

For clinical adoption, it is essential to understand why a particular PRS is high or low.

## B. Research and Assessment Questions

In this work we identified the following questions:

- How can researchers embed trustworthy AI principles into the AI design?
- What are the possible benefits and risks of using such an AI system?
- How can trustworthiness be embedded across the entire span of the design and implementation process of the XPRS system design and implementation?
- How can a Trustworthy AI assessment be performed together with a Human Rights Assessment?
- What lessons learned can help researchers working in similar areas?

## C. Frameworks for Trustworthy AI and Human Rights

In this case we used a co-design approach to assess the trustworthiness of the XPRS system by considering the legal, ethical, and technical robustness aspects of the AI system at the early design stage and for its potential implementation and use by medical professionals.

In this work, we considered the ethics guidelines for trustworthy artificial intelligence defined by the EU High-Level Expert Group on AI, where Trustworthy AI [20] is defined as:

(1) **lawful** - respecting all applicable laws and regulations

(2) **ethical** - respecting ethical principles and values
(3) **robust** - both from a technical perspective, considering its social environment

This general framework for Trustworthy AI (not specific to healthcare) is based on four ethical principles, rooted in fundamental human rights [20], namely:

1) respect for human autonomy

2) prevention of harm

3) fairness

4) explicability

and defines seven requirements for their operationalization, namely:

---

[1] *Z-Inspection® use-case main document.*

1) human agency and oversight
2) technical robustness and safety
3) privacy and data governance
4) transparency
5) diversity, non-discrimination, and fairness
6) societal and environmental well-being
7) accountability

Each requirement has a number of sub-requirements, as indicated in Table 1 below.

| **1. Human Agency and Oversight** |
| --- |
| *1.1 Fundamental Rights: Ensuring AI systems do not infringe upon human rights.* |
| *1.2 Human Agency: Empowering individuals to make informed decisions and maintain control over AI systems.* |
| *1.3 Human Oversight: Implementing mechanisms such as human-in-the-loop, human-on-the-loop, and human-in-command to oversee AI operations.* |
| **2. Technical Robustness and Safety** |
| *2.1 Resilience to Attack and Security: Protecting AI systems against adversarial attacks and malicious exploitation. Safeguarding AI systems against unauthorized access and data breaches.* |
| *2.2 Fallback Plan and General Safety: Establishing contingency plans for AI system failures to prevent harm.* |
| *2.3 Accuracy: Maintaining precise and correct outputs from AI systems.* |
| *2.4 Reliability and Reproducibility: Ensuring consistent performance and the ability to reproduce outcomes under varying conditions.* |
| **3. Privacy and Data Governance** |
| *3.1 Privacy and Data Protection: Complying with data protection regulations to ensure user privacy.* |
| *3.2 Quality and Integrity of Data: Utilizing accurate and relevant data to train and operate AI systems.* |
| *3.3 Access to Data: Regulating who can access data and under what conditions.* |
| **4. Transparency** |
| *4.1 Traceability: Documenting the AI system's development processes and decision-making pathways.* |
| *4.2 Explainability: Providing understandable explanations of AI system decisions to users and stakeholders.* |
| *4.3 Communication: Clearly conveying the capabilities and limitations of AI systems to users.* |
| **5. Diversity, Non-Discrimination, and Fairness** |
| *5.1 Avoidance of Unfair Bias: Implementing measures to detect and mitigate biases within AI systems.* |
| *5.2 Accessibility and Universal Design: Designing AI systems to be usable by diverse groups, including those with disabilities.* |
| *5.3 Stakeholder Participation: Engaging a broad range of stakeholders in the AI system development process to ensure inclusiveness.* |
| **6. Societal and Environmental Well-Being** |
| *6.1 Sustainability and Environmentally Friendly AI: Assessing and minimizing the ecological footprint of AI systems.* |
| *6.2 Social Impact: Evaluating the broader societal implications of AI deployment.* |
| *6.3 Society and Democracy: The impact of an AI system should be assessed not only on individuals but also on society, democracy, and institutions, especially in political and electoral contexts.* |
| **7. Accountability** |
| *7.1 Auditability: Enabling the AI system to be audited by independent parties.* |
| *7.2 Minimization and Reporting of Negative Impact: Establishing protocols to reduce and report any adverse effects caused by AI systems.* |
| *7.3 Trade-offs: When tensions arise between AI requirements, trade-offs must be explicitly acknowledged, evaluated for ethical risks, documented, and continuously reviewed to ensure accountability and necessary adjustments* |
| *7.3 Redress: Providing mechanisms for individuals to contest and seek remedies for decisions made by AI systems.* |

*Table 1. Requirements and sub-requirements Trustworthy AI. Reproduced from AI HLEG, 2019 [20]*

The EU trustworthy criteria cover three pillars (lawful, ethical, and robust), all of which are relevant to this use case and were adopted during the co-design process.

In this work, we applied Z-Inspection® [19], a holistic process for evaluating the trustworthiness of AI-based technologies across different stages of the AI lifecycle [14], [20].

For human rights, we considered the Council of Europe's HUDERIA methodology [26], which provides guidance and a structured approach to carry out risk and impact assessments for AI systems, and international human rights standards applicable in Korea [32]

## D. Contribution and Paper Structure

In this work, we have provided:

- A clear methodology to embed trustworthy AI principles into early AI design.

- A detailed analysis to identify possible benefits and risks of using [An AI system that explains PRS scores].

- Reflections and outline of potentially applicable domain-specific legislation (e.g., Korean data protection law, [Medical Service Act, Medical Devices Act,] etc.).

- Reflections and assessment of applicable AI-specific legislation (e.g., [Basic Act on the Development of Artificial Intelligence and the Establishment of a Foundation for Trustworthiness (or the AI Basic Act of Korea)]).

- Inclusion of a *human rights assessment*, using the Council of Europe's HUDERIA and International Human Rights standards applicable in Korea.

- A comparison of HUDERIA with the Z-Inspection® process.

- Lessons learned that can help researchers with similar research work.

The paper is structured as follows:

In Section II, the overall methodology based on the Z-Inspection® is described. Section III describes the co-design process in detail. Possible uses of the AI Systems are analyzed from different viewpoints, mapped to the Z-Inspection® framework for trustworthy AI. Section IV presents related work. Section V presents conclusions and the lessons learned.

## II. METHODOLOGY

## A. Z-Inspection® Process

In this use case, the Z-inspection® process was used as an ethically aligned co-design methodology [21], which helped the stakeholders assess their initial design decisions against the EU Framework for Trustworthy AI. Overall, the core of the co-design process is a cross-disciplinary conversation.

Z-Inspection® uses a *holistic* approach rather than monolithic, static ethical checklists.

This is of great value when analyzing use cases, as it mirrors the complexity of an AI system's use context. To be successful, the system must adequately meet technical, legal, ethical, and social requirements. Using a holistic process means that the Z-Inspection® assessment brings these different considerations together to provide an assessment of the system not only in different respects (technical, legal, ethical, social, etc.), but as a whole functioning socio-technical unit [19].

Another strength of this holistic approach is that it is dynamic. A standard checklist does not adapt to the case at hand. The holistic approach on the other hand determines which issues are central for the use case at different stages of the process and moves back and forth between intra- and inter-disciplinary discussions of which aspects of the case are most significant.

In this way, the process has a certain degree of plasticity, which means that its assessments will be tailored to the use-case at hand. It also means that the assumptions that guide the AI system's creation and deployment-as well as the assumptions of those researchers conducting the Z-Inspection®-are exposed and evaluated during the inspection. Finally, the dynamic lens reflects the commitment to an ongoing and iterative investigation of the harms and benefits of a particular AI system [19].

Unlike prior Z-Inspection® co-design use cases, this project also includes a human rights assessment using the Council of Europe's HUDERIA tool and international human rights standards applicable in Korea.

With a focus on trustworthiness, two pillars of the EU High-Level Expert Group's [20] guidelines for trustworthy AI were considered: lawfulness and ethical issues. The focus on trustworthiness is confirmed relevant to the Korean context, as in May 2021, the Ministry of science and ICT adopted a strategy as a follow-up to its ethics guideline (Dec 2020). The strategy, best translated as "Strategies to promote Trustworthy AI for Human-centered AI", and trustworthy AI have been at the center of government-wide policy discussion in Korea.

Like the EU Guidelines, Z-Inspection® does not address the lawful aspects of trustworthy AI in detail. Nevertheless, considering the importance of the legal aspect for trustworthiness, the emerging AI-specific legislation, and considering the importance of human rights respecting AI systems and use, the Z-Inspection® methodology was adapted in this use case to include:

> 1) reflections and outline of potentially applicable domain-specific legislation (e.g., Korean data protection laws, Medical Service Act, Medical Devices Act, etc.).

> 2) reflections and assessment of applicable AI-specific legislation (e.g., the AI Basic Act of Korea)

> 3) the inclusion of a human rights assessment, using the Council of Europe's HUDERIA tool and International Human Rights standards applicable in Korea.

The legal assessment is not aimed at ensuring legal compliance, and thus, should not be considered specific legal advice.

### Trustworthy AI Co-design

*Co-design is defined as a collective creativity, engaging end-users and other relevant stakeholders, that applies across the entire span of the design process [21].*

In the rest of the paper, we use the term stakeholders to denote the actors who have direct ownership of the development and deployment of the AI system [19].

Fig. 1 below illustrates the Z-inspection® Co-design Process. Our work started with an early-stage design and prototype, considered the initial design as a claim to be assessed using socio-technical scenarios and an evidence based approach, and provided feedback to the stakeholders of issues and possible risk and mitigations to be considered for future changes in the design of the AI system (ref. Redefinition of the AIs design´s goal/purpose in Figure 1), and in addition serving as a reference model for future similar projects.

The various co-design tasks for this use case are detailed in the remainder of this report.

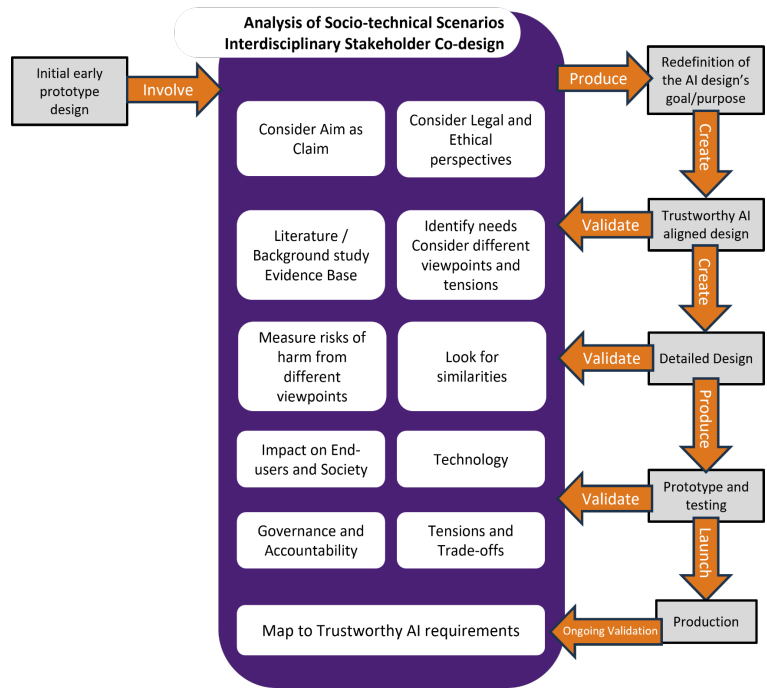

*Figure 1. Z-Inspection® Co-design Process, adapted from [13]. The figure illustrates the iterative stages of the Z-Inspection® co-design methodology applied in this use case.*

## Sociotechnical Scenarios

Sociotechnical scenarios are essential in the Z-Inspection® process, linking the technical capabilities of AI systems with their societal impacts. They uncover potential ethical, technical, and legal issues while fostering collaboration among stakeholders.

By collecting relevant resources, a team of interdisciplinary experts creates sociotechnical scenarios and analyzes them to describe the aim of the AI system, the actors and their expectations and interactions, the process where the AI system is used, and the technology [19].

By simulating real-world contexts, these scenarios identify unforeseen risks and tensions, ensuring AI systems are transparent, accountable, and aligned with societal values. They also clarify human-machine roles and stakeholder expectations, helping to evaluate the broader implications of AI deployment.

We found it useful to collect a written summary of the meetings among the stakeholders and the Z-Inspection® experts, where the information is organized according to the structure presented in Appendix A.

### Claims, Arguments, and Evidence

An important part of our assessment process is to build an evidence-base through the sociotechnical scenarios to identify tensions as potential ethical issues to be discussed further.

Claims for technological capability (for example aim, performance, architecture, or functionality) serve as an important input in developing the evidence base. This is an iterative process among experts of the assessment team with different skills and backgrounds with a goal to understand technological capabilities and limitations [19].

This part of the co-design process consists of reviewing (documents, codes, designs, etc.) and creating an evidence base to verify and support any claims made by producers of the AI system and other relevant stakeholders. This task is domain specific.

## B. Creation of Interdisciplinary Team

When working in co-creation, managing different viewpoints between experts composing the assessment team is an essential part of the process. There may be tensions when considering what the relevant existing evidence to support a claim is. During the co-creation process, experts in our team included experts in public health and healthcare, specialists in genomics, and evidence-based diagnosis, ethics, law, and AI. For managing different viewpoints, the process requires researchers to bring their own ideas and arguments to the discussion of aspects of the case while at the same time understanding that their input is a contribution to teamwork, not a matter of "winning the argument." [19].

## C. Split the Work in Working Groups

This step of the co-design phase consists of identifying possible ethical, technical, and legal issues for the use of the AI within the given boundaries and context.

To do so, we have been working with three parallel Working Groups (WGs), one focused on *ethics, law and human rights* (WG1), one focused on *healthcare* (WG2) and one focused on *AI technology* (WG3). Each WG analyzed the sociotechnical scenarios and produced preliminary reports, working independently and in parallel to avoid cognitive biases and to take advantage of their unique perspective and expertise.

## D. Creation of Reports

Preliminary reports were shared with the entire team for feedback and comments. These interdisciplinary interactions among experts with different backgrounds allow each WG to consider the viewpoints of other experts when delivering their final reports.

Each final report is then written using free text and an open vocabulary to describe the possible risks and issues found when analyzing the AI system. For example, each WG report may list the identified ethical, technical, domain-specific (i.e. medical) and legal issues described using an open vocabulary.

## E. Mappings to the Framework of Trustworthy AI

In evaluating the use case, the team of experts used a process that serves to unify the assessment process and allows agreement between the various experts with different backgrounds. The process began with describing tensions between ethical values using open vocabulary and gradually narrowing the options down to finally agree on a closed vocabulary description. The closed vocabulary was from the EU framework.

## F. Consolidation Process of Mapping

The "issues" described in free text are then mapped by each WG using templates (called rubrics) to some of the four ethical principles and the seven requirements defined in the EU framework for trustworthy AI [20].

With this mapping, the reports are developed from an open vocabulary to a closed vocabulary (i.e., the templates). We call this process "*mapping*''. While working independently, each working group adopted different/similar strategies to perform the *mappings*.

The *mappings* and the common vocabulary are especially important in the interdisciplinary context that any technological assessment necessarily implies. The *mappings* required experts to translate their own disciplinary methods and cultural perspectives into a single language that everyone understands.

## G. Give Recommendations

Depending on the list of identified ethical issues, we give recommendations with the aim of helping to improve the design of the AI. Developers/researchers could use the feedback and results from the co-design to improve the AI's design and better align it with the ethical/legal, technical, and domain-specific challenges.

## III. TRUSTWORTHY CO-DESIGN AI

Different stakeholders are affected differently by the AI system and may hold different values. Resolving ethical tensions requires us to understand diverse public opinions on questions related to these tensions and inevitable trade-offs. In the co-design process, this is considered during the analysis of socio-technical scenarios where we identify the various stakeholders and agree on ways to involve them in the design process.

## Analyzing of the Sociotechnical Scenarios from Different Viewpoints and Mapping to the Framework for Trustworthy AI

The Assessment Phase of the process began with the creation of sociotechnical scenarios.

As part of the Z-Inspection® process, Working Group 1 (WG1)[2] identified and considered ethical and legal (including human rights) issues associated with an Interpretable and Explainable Polygenic Risk Score (XPRS) AI Tool to Predict Type 2 Diabetes using genome data, including issues arising

---

[2]The link to the socio-technical shared doc: [https://docs.google.com/document/d/1hgbuy-AAzUfnJS9yAyIQIiakBs3wh8I2ApXK0rzSTKw/edit?tab=t.0]

WG1: Ralf Beuthan, Elisabeth Hildt (co-lead), Heejin Kim, Sira Maliphol, Vadim Pak, Stephan Sonnenberg, Emilie Wiinblad Mathez (co-lead).

from the use of PRS itself. Working Group 2 (WG2)[3] and Working Group 3 (WG3)[4] identified and considered domain-specific and technical issues.

In the following, we present the results of the work of the three parallel Working Groups (WGs), one focused on ethics, law, and human rights (WG1), one focused on healthcare (WG2) and one focused on AI technology (WG3).

## Issues raised by the Technical and Medical Working Groups

WG2 and WG3 have provided full reports with identification of technical and medical issues.

Below is a short summary of three main themes resulting from the work of WG2 and WG3. The full list of issues is presented later in this section.

**Accuracy in predictions**: Both working groups noted the need for accuracy of PRS predictions, which would require being validated in real-world settings for clinical utility, as well as the need for warning if the test data falls outside the range of the entire training population. This is particularly relevant if/when used for individuals with ancestry other than Korean or Japanese.

**Interpretability**: The XPRS includes visual SNP and gene contributions which are important as low interpretability limits clinical trust and regulatory acceptance. Questions are raised, however, regarding whether clinicians would be able to better explain PRS using the XPRS, as this has not been tested in practice or received user feedback. It was suggested that the explanation needs to be tailored to the doctors/researchers receiving the explanation making them understandable, interpretable with properly highlighting the SNP, genes, and their contributions and confidence level. If visualizations are unclear or hard to interpret, clinicians may misinform patients. Patient and clinician feedback will be important.

**Benefits and "do no harm"**: Whether patients will benefit from prevention suggestions and lifestyle advice depends on the real-life scenarios, as well as the evidence linked that knowledge of risk will trigger behavior change and early interventions in the manner intended.

The Technical Working Group has assessed that the AI tool is at 5-6 Technology Readiness Level and therefore not yet ready for clinical use [22].

## Identified Medical Issues and Tensions (WG2)

From a medical perspective, WG2 examined whether the proposed PRS AI system can support clinical reasoning and clinician-patient communication in a way that is clinically meaningful, without creating new risks for patients or clinicians. The assessment therefore went beyond predictive performance and focused on how medical meaning, uncertainty, and clinical context are expressed and communicated through the system.

WG2 developed its findings through an iterative, discussion-driven process across a series of meetings. In an initial phase, several sessions included the development team and focused on clarifying the intended clinical purpose, the current design choices, key assumptions about data and workflow, and what "interpretability" and "explainability" should mean in real clinical encounters. In a later phase, the working group continued in assessment-only meetings, where the discussions were synthesized and consolidated into a set of medically relevant issues. The group treated these issues as an evidence-linked summary of recurring clinical tensions and design considerations, intended to inform further co-design and refinement rather than to serve as final clinical recommendations.

---

[3]WG2: Megan Coffee (co-lead), Na Yeon Kim, Pedro Kringen(co-lead), Haekyung Lee, Seunggeun Lee.

[4]WG3: Jesmin Jahan Tithi(co-lead) , Magnus Westerlund(co-lead). Project Lead and Coordination: Roberto V. Zicari.

In this work, WG2 assessed the system along clinically relevant dimensions of trustworthiness, including the clarity of clinical utility, the interpretability of genetic risk information for decision-making and communication, the fit with real-world clinical workflows, and the risk of inappropriate reliance on outputs that have not yet been sufficiently validated for use in practice.

Particular attention was given to whether the system's outputs can realistically support explainability in the clinician-patient interaction, which is a central stated aim of the use case.

Given the early development stage of the system (TRL 5-6), the medical assessment did not aim to produce definitive clinical recommendations. Instead, it identified medically relevant issues, tensions, and design considerations that would need to be addressed before safe and responsible clinical deployment.

Building on this framing, WG2 approached trustworthiness as a clinical question of purpose, interpretation, and risk under uncertainty, and grouped the identified concerns into a small number of recurring themes.

A first recurring concern is the system's *clinical utility*. The report describes an interpretable and explainable PRS tool, but the current version still leaves the clinical role of the output underspecified. Clinicians need a clear statement of intended use that answers practical questions: who uses the tool, at what point in a care pathway, and for which decisions. If the system targets screening, risk communication, triage, follow-up intensity, or lifestyle intervention counselling, the interface and the explanation strategy need to reflect that specific task. Without that framing, the tool risks becoming an attractive visualization without a defined decision context. This gap also makes it hard to evaluate benefit versus burden, because "benefit" depends on a use case and an action. The co-design aim, therefore, depends on a sharper description of clinical use scenarios, including who takes responsibility for interpretation and how the output integrates with existing risk factors and guidelines.

A second cluster concerns i*nterpretability and explainability* as experienced by clinicians and patients. The tool offers gene- and SNP-level views, but genetic explanations easily become technically correct while remaining clinically unusable. A clinician needs an explanation that supports judgement under time pressure, not a lecture in statistical genetics. A patient needs an explanation that supports understanding and autonomy, not a dense display that increases anxiety or produces false reassurance. In this sense, "explainable" does not mean "more information". It means the system selects and organizes information to match the user's role, health literacy, and decision needs.

The WG2 discussion, therefore, returned repeatedly to explanation design as a communication problem, not only a modelling problem. The system needs a clear separation between professional-facing explanations and patient-facing explanations, with an explicit account of what each user group should take away from the output. This matters directly for the stated aim of improving clinician-patient communication, because communication fails when the explanation is inappropriate for the person receiving it.

A third theme concerns *clinical safety in the presence of uncertainty.* PRS outputs are probabilistic and population derived. They can inform prevention strategies, but they also invite overinterpretation when users treat a risk score as a diagnosis, a prognosis, or a direct trigger for action.

The WG2 work, therefore, highlighted the risk of over-reliance on predictions that have not yet been validated for the deployment context, and the related risk of automation bias. If the interface presents a single risk estimate without clear uncertainty cues, guardrails, and guidance on what the score should not be used for, the system can shift clinical judgement in unintended ways. This is not an abstract concern. Risk scores can change patient behavior, affect clinical attention, and alter how clinicians prioritize follow-up. The report, thus, treats "do no harm" in practical terms: the system needs to support informed human judgement, not replace it. The design should help users understand the limits of the score, the conditions under which the score becomes unreliable, and the fact that risk does not translate automatically into clinical action.

A fourth matter relates to *external validity and fairness across populations*. PRS models often show performance differences when they move across ancestry groups and clinical contexts. The current tool has been developed using East Asian biobank resources, which creates a clear boundary around who the model most likely serves well. Even within broadly defined populations, genetic variation and environmental context introduce heterogeneity that the system must acknowledge.

The medical concern here is twofold. First, uneven validity can create unequal clinical benefit, and in the worst case it can mislead precisely those groups already underserved by precision medicine. Second, lack of explicit population caveats invites silent misapplication once the tool spreads beyond its original setting, for example through research collaborations, commercial interest, or clinical enthusiasm. The system therefore needs explicit statements about population scope, subgroup performance evidence, and clear rules for when clinicians should treat an output as unreliable. These are not purely technical points. They determine whether clinicians can use the tool responsibly and whether patients can trust the meaning of the results.

A fifth theme concerns *selection bias* and the difference between *research cohorts and real-world patients.* Biobank populations often differ from clinical populations in systematic ways, including access to care, health-seeking behavior, and selection effects tied to recruitment. When the model learns patterns from such datasets, it risks overfitting to a research context and underperforming in routine care. The clinical implication is not only reduced accuracy. It is also a shift in who benefits, because those who resemble the training population receive better calibrated results than those who do not. Hence, WG2 treated replication, out-of-sample testing, and careful documentation as clinical trustworthiness prerequisites, not optional technical refinements. Clinicians cannot interpret a risk score responsibly if the system cannot demonstrate robustness under the variability of real clinical use.

Finally, WG2 raised a recurring concern about *proportionality in genomic data use.* The system operates on highly sensitive genetic information. If the system processes more genomic data than necessary to achieve its stated purpose, it increases privacy risk without a clear clinical gain. This matters even in early development, because data pipelines, storage patterns, and sharing practices tend to persist once established. The report therefore frames data minimization as part of medical trustworthiness, not only as a legal requirement. A system that aims to improve prevention and communication should also limit exposure to avoidable harms that follow from unnecessary data collection, retention, or reuse.

Across the WG2 findings, several tensions recur. Improving transparency and explainability tends to increase interface complexity, which can reduce usability in real clinical workflows. Moving fast to prototype novel explainability features conflicts with the need for careful validation and appropriate constraints on clinical use. Finally, improving predictive accuracy by using richer genomic information must be balanced against proportionality and data minimization. These tensions are not unique to this implementation, but typical for PRS-based clinical support tools, and they shape what "responsible explainability" can realistically look like at the current maturity level.

Taken together, these themes explain why the WG2 issues connect directly to core trustworthiness concerns, including transparency and meaningful explanations, human agency and oversight in clinical decision-making, prevention of harm under uncertainty, fairness and population validity, and privacy and data governance. Several of these issues are not unique to this specific implementation but reflect recurring clinical challenges for PRS-based decision support more broadly, particularly around clinical utility, communication of probabilistic risk, and appropriate use under uncertainty.

This point is central to the stated aim of improving explainability in the clinician-patient interaction. The system's emphasis on interpretation and visualization of PRS outputs provides a plausible pathway to support better clinician-patient communication, but WG2 found that explainability remains clinically fragile at the current maturity level. Gene- and SNP-level explanations can be technically correct while still difficult to use in real consultations, and they can invite misinterpretation if PRS outputs are perceived as deterministic or directly actionable.

Thus, WG2 treated explainability as a clinical communication task rather than a simple disclosure of model internals. In practical terms, this would require layered explanations adapted to user and context, with a clear separation between clinician-facing support for judgement and patient-facing communication, and with explicit communication of uncertainty, limitations, and intended-use conditions. Where relevant, genetic explanations should also be anchored in clinical context, including conventional risk factors and modifiable contributors, to support meaningful dialogue and reduce the risk of over-reliance.

In terms of what changed through the co-design work, the discussions reinforced that explanation mechanisms need to be treated as first-class design objectives rather than add-ons. At the same time, WG2 noted that the credibility of any claim that the system improves clinician-patient explainability ultimately depends on validation in realistic clinical settings. A key tension remains that richer explanatory detail can reduce usability and clarity in time-constrained consultations, yet simplification can improve usability while risking oversimplification.

The assessment does not aim to finalize clinical recommendations at this stage. It documents the clinical questions and design tensions that must be resolved through further co-design and validation before the tool can credibly support patient care at scale. These recurring themes are reflected in the specific medical issues, tensions, and mitigations summarized in Table 2.

Table 2 *Medical Issues, Tensions Mappings, and Mitigations Strategies. Mapping numbers refers to HLEG requirements and sub-requirements [20] [High-Level Expert Group on Artificial Intelligence. Ethics guidelines for trustworthy AI. European Commission]).*

| Medical Issues # | Ethical Issue / 7 Requirements of Trustworthy AI/HLEG Mapping | Tensions |
|---|---|---|
| Assessment restricted to the current TRL level. | | HLEG mapping |

| #1: Unclear clinical utility of the PRS tool | Transparency & Human Agency and Oversight | **Human Agency and Oversight**: Accessibility and Universal Design, Utility |
|---|---|---|
| **Description** | | vs. |
| It's not explicitly defined how the PRS output supports actual medical decision-making. | | **Technical Robustness and Safety:** More safeguards and checks can add interface and workflow complexity (if UI gets denser). |
| **Why is this concerning?** | | ____________ |
| If it is not clear how the PRS result is supportive of real clinical decisions, clinicians may interpret and use the results inconsistently, risking that this dilutes the quality of patient care. Clinically well-defined questions and a clear rationale for changes in clinical practice or patient management are needed. | | **Human Agency and Oversight:** Empowering individuals to make informed decisions and maintain control over AI systems |
| | | vs. |
| **Suggested solution and/or mitigation strategy** | | **Technical Robustness and Safety** and security |
| Define potential clinical use and present different use cases. Refer to medical personnel, including endocrinologists and primary care physicians. | | ____________ |
| | | **Transparency:** Explicitly stating what the PRS can and cannot support clinically is required for correct use. |
| | | vs. |
| | | **Human Agency and Oversight**: If the clinical role remains vague, clinicians may still over- or under-rely on the tool. |
| | | ________________ |
| | | **Diversity, Non-Discrimination, and Fairness**: The PRS tool must work across diverse clinical settings and user groups, not only in specialized centers. |
| | | vs. |

| | | **Accountability**: Stronger governance and regulatory controls can slow adaptation to local workflows and needs. |
|---|---|---|
| **#2: Limited interpretability of genetic explanations**<br><br>**Description**<br><br>Concerns that gene/SNP-level visualizations are complex and may not be usable by clinicians without training.<br><br>**Why is this concerning?**<br><br>If gene/SNP visualizations are too technical or not explained, clinicians may misinterpret the output risk or misinform a patient about it.<br><br>**Suggested solution and/or mitigation strategy**<br><br>Continue to incorporate research from this ongoing field of research and incorporate multivariable analysis with other important risk factors (body mass index, smoking, diet, exercise, medications). | Transparency & Technical Robustness and Safety | **Societal and Environmental Well-Being:** Introducing a complex PRS–XPRS tool must justify its resource, training, and workflow burden at population level<br><br>vs.<br><br>**Accountability**<br><br>The system owner must implement controls to reduce foreseeable harms and maintain a post-deployment process to detect, document, report, and correct adverse effects in real-world use.<br><br>———————————<br><br>**Human Agency and Oversight**<br>Must show clinicians can use it<br><br>vs<br><br>**Transparency**<br><br>Multivariable context improves sense-making. |

| #3: Potential over-reliance on unvalidated predictions | Diversity, Non-Discrimination, and Fairness & Human Agency and Oversight & Transparency, & Accountability: | **Human Agency and Oversight**: Clinicians need a simple, clearly framed PRS output they can safely act on in routine care. |
|---|---|---|
| **Description** | | vs |
| Risk scores may be acted upon without sufficient awareness of their uncertainties or limitations, and combining behavioral, environmental, and genetic risks for one individual is difficult. | | **Technical Robustness and Safety**: More sophisticated models and uncertainty layers increase complexity and can be harder to use correctly in real workflows. |
| **Why is this concerning?** | | ________________ |
| Without awareness of their uncertainty or limitations, responses may not account for the full picture. | | **Human Agency and Oversight:** Clinicians must remain free to override or ignore PRS outputs when they conflict with clinical judgement. |
| **Suggested solution and/or mitigation strategy** | | vs. |
| Developers should design the system to support informed human judgment, not to be a "black box oracle". | | **Accountability:** If clinicians over-rely on the score, the responsibility and auditability of decisions become blurred. |
| Uncertainty quantification | | ________________ |
| Integrate mechanisms that communicate prediction confidence or uncertainty (e.g., probabilistic outputs, calibrated risk scores) instead of binary "yes/no" outputs | | **Human Agency and Oversight:** Directly targets automation-bias risk in this PRS decision support setting |
| | | vs. |
| | | **Transparency**: Calibrated risk and explicit uncertainty communication are required so users recognize limits and do not treat predictions as validated facts. |
| #4: Ethical concern: excessive genomic data use | Privacy and Data Governance | **Technical Robustness and Safety**: Adding more genomic variants may marginally improve PRS accuracy for some patients. vs. |
| **Description** | | **Privacy and Data Governance:** Data-minimization requires using only the smallest genomic |
| Raises issue of data minimization and whether the system uses more genomic data than necessary. | | |
| **Why is this concerning?**<br><br>Using full genomes when a fraction would suffice raises ethical issues concerning data minimalism, privacy, and proportionality.<br><br>**Suggested solution and/or mitigation strategy**<br><br>Explain why this amount of data is needed. Is it needed? Perform calculations for different quantities of genetic data/SNPs and determine whether large quantities yield substantially more accurate predictions than smaller quantities. | | subset that passes proportionality and accuracy tests. |
| **#5: Algorithmic Bias Across Populations**<br><br>**Description**<br><br>Model calibrated only on East Asian biobanks.<br><br>**Why is this concerning?**<br><br>Doesn't generalize to underserved populations or real-world variability. Even if the population is quite homogenous, not all genes will be equally represented in any sample population.<br><br>**Suggested solution and/or mitigations strategy**<br><br>Propensity scoring patients vs similar groups? Larger confidence intervals/margin of errors. If individuals come from very specific isolated, native populations, may list caveats about the performance of the tool. | Diversity, Non-Discrimination, and Fairness & Transparency | **Diversity, Non-Discrimination, and Fairness:** Requires explicit subgroup evaluation and labelling so users see that performance differs across populations.<br><br>vs.<br><br>**Accountability**: Sensible interim guardrails are needed so the East Asian-calibrated PRS is not used as if it were universally valid.<br><br>________________ |

| #6 Overfitting to Research Data | Technical Robustness and Safety & Diversity, Non-Discrimination, and Fairness | **Technical Robustness and Safety:** Rapidly iterating on biobank data can produce seemingly strong PRS performance and encourage early deployment. |
|---|---|---|
| **Description** | | |
| Training on biobank populations with regular health access. | | vs. |
| | | **Accountability**: Regulatory safeguards must delay clinical use until external validation shows the model works beyond well-served biobank populations. |
| **Why is this concerning?** | | |
| Doesn't generalize to underserved populations or real-world variability. Even if the population is quite homogenous, not all genes will be equally represented in any sample population. | | |
| | | ———————— |
| | | **Technical Robustness and Safety:** Canonical anti-overfitting tools and explicit replication can stabilize performance across datasets. |
| **Suggested solution and/or mitigations strategy** | | vs. |
| When possible, use larger and more diverse datasets that better represent the intended deployment context. Apply cross-validation and out-of-sample testing to verify generalizability. Incorporate regularization techniques to reduce model complexity. Conduct replication studies with independent datasets. Clearly document dataset limitations and communicate uncertainty to stakeholders. | | **Transparency:** Detailed documentation of datasets, model limits, and replication results is needed so clinicians and regulators can judge when the tool is safe to use. |

## Lessons learned WG 2 Medical

Building on the issues and tensions described above, and the structured mapping in Table 2, the WG2 work yielded several lessons for future development and assessment of XPRS.

The analysis confirmed that clinical utility cannot be inferred from interpretability alone. The themes on unclear intended use and decision context show that XPRS will only become clinically meaningful once concrete use cases, responsibility allocations, and integration points in existing pathways are explicitly defined and tested. This implies that future iterations should treat specification of purpose and workflow fit as primary design tasks, on par with improving model performance or adding new visualization features.

The findings on interpretability and communication underscore that explainability must be designed as a clinical communication function, not just exposure of model internals. The distinction between clinician-facing and patient-facing explanations, outlined in the main text and reflected in the table entries on limited interpretability, needs to be operationalized through layered interfaces, multivariable context, and explicit uncertainty cues. WG2 learned that these explanation mechanisms should be treated as first-class design objectives and validated empirically in realistic consultations before strong claims about improved communication are made.

The systematic concern about over-reliance on unvalidated predictions and automation bias suggests that any clinical deployment must position XPRS explicitly as decision support. The table entries on "over-reliance on unvalidated predictions" and "overfitting to research data" indicate that guardrails, calibrated risk presentation, and clear "do not use for" statements are as important for safety as the numerical accuracy of the score itself. WG2 therefore concludes that risk communication, usage constraints, and monitoring for misuse should be built into both the interface and the governance surrounding the tool.

Also, the themes on external validity, cohort bias, and selection effects confirm that population scope is a core part of clinical trustworthiness. Because the current model is trained on East Asian biobank cohorts, WG2 learned that explicit subgroup performance reporting, population caveats, and replication in routine-care settings are prerequisites for responsible scaling, rather than follow-up luxuries. The mapping in Table 2 to diversity, fairness, and technical robustness requirements highlights that questions of "who the tool serves well" and "for whom it is unreliable" must be answered before national or cross-population deployment is considered.

Finally, the work on proportionality and genomic data volume shows that data-minimization is also a medical design question. The identified issue on excessive genomic data use and its associated tensions indicate that the quantity of genetic information should be justified against clinically meaningful gains in calibration and discrimination, using ablation and sensitivity analyses rather than defaulting to maximal panels. Accordingly, WG2 treats proportionality in genomic data use as part of clinical trustworthiness and recommends that future development link features-set choices directly to evidence about added value for concrete clinical decisions.

## Identified Technical Issues and Tensions (WG3)

From a technical perspective, WG3 examined whether the proposed PRS-based AI system is technically robust, interpretable, and maintainable in a way that can credibly support its intended clinical use, without introducing hidden failure modes or operational risks. The assessment went beyond model accuracy in isolation and focused on how explanation mechanisms, uncertainty handling, monitoring, and lifecycle management are implemented at the system level.

WG3's discussions focused on how technical design choices relate to requirements such as transparency, robustness, accountability, and human oversight. These identified issues are intended to inform further system refinement and governance planning rather than to serve as final technical prescriptions.

Given the early maturity of the system (TRL 5-6), WG3 approached trustworthiness as an emergent property of the full technical pipeline, including model behavior, interface design, monitoring mechanisms, and maintenance practices. Consequently, the assessment focused on whether the system can reliably communicate its limits, degrade safely under uncertainty, and remain technically dependable over time.

A first recurring technical concern relates to *tailored explanation mechanisms for different user roles*. The system provides SNP- and gene-level contribution information intended for clinicians and researchers, under the assumption that patients will not directly interact with the raw outputs. From a technical standpoint, this raises challenges around explanation granularity, abstraction, and audience alignment. Explanations that are sufficiently detailed for expert users' risk being incomprehensible or misleading if surfaced to non-expert users, while overly simplified views may obscure clinically relevant uncertainty or model limitations. WG3 identified a core tension between transparency and comprehensibility: explanation pipelines must support role-specific views without duplicating logic or introducing inconsistency. This requires explicit technical separation of explanation layers and clear assumptions about who the explanation is designed for and how it may be reused or exposed downstream.

A second cluster of issues concerns *confidence estimation and out-of-distribution (OOD) detection*. WG3 identified confidence reporting and OOD warnings as essential technical safeguards for robustness and safety. However, these mechanisms are only effective if they are well-calibrated, clearly defined, and consistently presented. Poorly calibrated confidence scores may be ignored or misinterpreted, while overly sensitive OOD detection can generate excessive warnings that lose practical meaning. From a systems perspective, this creates a tension between accuracy signaling and usability. WG3 also highlighted the need for clear technical definitions of what constitutes "outside training coverage" and how such conditions should constrain downstream use, logging, and monitoring.

A third theme concerns *feedback mechanisms for clinicians*. Technically, enabling user feedback creates opportunities for continuous improvement, error detection, and post-deployment learning. At the same time, feedback systems introduce challenges related to data quality, bias, and accountability. Feedback may be sparse, inconsistent, or systematically skewed toward certain environments or user groups, which can distort subsequent updates. WG3 identified a tension between empowering human oversight and maintaining reliable model governance. Without a clear technical pipeline for ingesting, validating, and acting on feedback, such mechanisms risk becoming symbolic rather than operational, while still creating expectations of responsiveness.

A fourth recurring concern relates to *accuracy thresholds and automated notifications*. WG3 assessed the proposal to monitor model performance and automatically notify stakeholders, or suspend use, if accuracy falls below a predefined medically acceptable threshold. From a technical standpoint, this requires reliable performance metrics, representative monitoring data, and clearly defined triggers. WG3 identified a tension between safety and continuity: conservative thresholds may lead to frequent interruptions, while permissive thresholds risk prolonged exposure to degraded performance. The group also noted that performance degradation can arise from multiple sources, including data drift, pipeline changes, or upstream dependencies, which complicates attribution and response. Robust implementation therefore requires not only thresholding but also diagnostic tooling and transparent reporting.

A fifth theme concerns *long-term maintenance and lifecycle management*. WG3 treated maintenance as a core technical requirement rather than a post-deployment concern. Genomic models are particularly sensitive to changes in population data, reference databases, and clinical interpretation standards. Without a clear maintenance plan, including versioning, retraining criteria, documentation updates, and dependency management, the system risks silent degradation. WG3 identified a tension between sustainability and resource constraints: maintaining technical robustness over time requires ongoing investment that may conflict with short-term development incentives. From a trustworthiness perspective, unclear ownership of maintenance responsibilities creates systemic risk.

Across these themes, WG3 identified several recurring technical tensions. Increasing transparency and safeguards often increases system complexity, which can reduce reliability and interpretability at the interface level. Adding monitoring, feedback, and alerting mechanisms improves safety but introduces operational overhead and new failure modes. Optimizing for performance and explainability simultaneously can expose trade-offs between model complexity and controllability.

Taken together (Table 3), the WG3 findings show how technical design choices directly shape the system's ability to meet core trustworthiness requirements, including transparency, robustness, accountability, and human oversight. Many of the identified issues are not unique to this implementation but reflect recurring challenges in deploying PRS-based decision-support systems in real-world settings.

WG3 therefore treats explainability, uncertainty handling, monitoring, and maintenance as first-class technical objectives rather than auxiliary features. At the current maturity level, the system demonstrates promising design intent, but its technical trustworthiness remains conditional on resolving the identified tensions through further engineering, validation, and governance alignment.

The assessment does not aim to finalize technical readiness for deployment. Instead, it documents the key technical risks and design considerations that must be addressed before the system can reliably support clinical use at scale.

Table 3 *Technical Issues and Tensions*

| Technical Issue | Ethical Issue / 7 Requirements of Trustworthy AI | Tensions |
|---|---|---|
| **#1: Tailored explanations for doctors vs. patients**<br><br>The explanation needs to be tailored to the doctors/researchers receiving the explanation making them understandable, interpretable with properly highlighting the SNP and gene and their contributions and confidence level. We assumed the patients only receive outcomes through the doctors and never see the direct output from the tool. If the patient also sees the output, output/explanation needs to be adjusted based on the understanding of the patients. It opens a can of worms! | *Transparency, Human agency & oversight, Diversity & non-discrimination* | - **Transparency vs. Comprehensibility:** Overly technical for patients vs. too simplified for doctors.<br><br>- **Autonomy vs. Paternalism:** Patients may want raw outputs, but direct exposure may harm rather than empower.<br><br>- **Fairness:** Different literacy/cultural backgrounds may lead to unequal understanding. |

| #2: Confidence levels & Out-of-Distribution (OOD) warnings<br><br>The tool should always show confidence in its prediction and if the test data falls outside the range of entire training population, it should show warning that the test data is completely outside the range of training sample coverage | *Transparency, Technical robustness & safety, Accountability* | - **Accuracy vs. Trust:** Confidence scores may be misinterpreted or ignored.<br><br>- **Non-maleficence vs.**<br><br>**Usability:** Too many warnings → alarm fatigue; too few → hidden risks.<br><br>- **Responsibility:** If warnings are ignored, is blame on the clinician or the tool developer? |
|---|---|---|
| #3: Feedback mechanism for doctors<br><br>The tool should enable the doctors to provide feedback if you want to and at a regular Cadence | *Human agency & oversight, Accountability, Diversity & non-discrimination* | - **Autonomy vs. Burden:** Feedback empowers clinicians but risks extra workload.<br><br>- **Accountability vs. Liability:** Feedback given but not acted on who is responsible?<br><br>- **Justice:** Without inclusive feedback, improvements may privilege elite settings. |
| #4: Accuracy thresholds & automatic notifications<br><br>If the tool's accuracy ever falls below a medically acceptable threshold it should send automatic notification to doctors/developers/maintainers/governing agencies about this and it should not be further used without restoring its accuracy again. | *Technical robustness & safety, Accountability, Transparency* | - **Beneficence vs. Non-maleficence:** Using a low-accuracy tool risks harm; suspending it may delay care.<br><br>- **Responsibility Distribution:** Who enforces suspension — developers, hospitals, regulators?<br><br>- **Trust vs. Transparency:** Sudden suspension may erode clinician trust if not explained clearly. |

| #5: Maintenance plan for the tool | *Accountability, Technical robustness & safety, Societal & environmental well-being* | - **Sustainability vs. Resource Constraints:** Maintenance requires ongoing investment, conflicting with profit motives.<br><br>- **Accountability vs. Ownership:** Who owns long-term responsibility — vendor, healthcare provider, regulator?<br><br>- **Justice:** Under-resourced hospitals may be more vulnerable to risks if maintenance lapses. |
|---|---|---|
| The developers should put together a plan of maintenance for the tool. | | |

## The legal/human rights and ethical framing and main findings

In this section we present the results of the work of WP1, including:

> 1) reflections and outline of potentially applicable ***domain specific legislation*** (e.g., Korean data protection laws, health care law, liability law, criminal law, discrimination law, etc.).
>
> 2) reflections and assessment of applicable ***AI specific legislation*** (e.g., AI Basic Act/Law of Korea; AI Framework Act of Korea)
>
> 3) the inclusion of a human rights assessment, using the Council of Europe's HUDERIA tool and International Human Rights standards applicable in Korea.

The legal assessment is not aimed at ensuring legal compliance, and thus, should not be considered specific legal advice.

### Reflections and outline of potentially applicable domain specific legislation

In addition to potentially applicable AI legislation in Korea, domain specific laws may apply. It is beyond the scope of this assessment to consider legal compliance with Korean law, however WG1 did suggest that health care laws, data protection, and the lawfulness of taking DNA samples of the population at birth, for the underlying PRS, might be among applicable laws to comply with.

There are three areas of discussion concerning rules and principles regulating various aspects of responsible development, use and application of the XPRS AI tool in the Korean context. First, it is essential to consider "National Guidelines for AI Ethics" prepared by the Ministry of Science and ICT (MIST) and adopted nationally in December 2020. Second, as Korea recently passed the "AI Basic Act" or the world's second AI-specific legislation in December 2024, it is also of importance to see how the current case of XPRS may be evaluated against this local law. The additional concerns lie in the legislations that are not AI-specific but still relevant to examine whether the XPRS case can be legally acceptable. Nevertheless, the main goal of our report is not concerned with discussing legal compliance in full. We rather provide a brief overview of rules and principles that are helpful for understanding the normative implications of developing and using the XPRS AI tool.

## Reflections and assessment of applicable AI specific legislation

### AI-related law

Two questions were considered.

1) Is the XPRS an AI system which would fall under the definition of AI in the Korean AI legislation? (or the EU AI Act Art 3) [23].

2) If so, would this system be classified as High-Risk by the EU AI Act and/or the Korean AI Law?

Regarding the first question, WG1 relied on the expertise of WG3, and it was found that the XPRS is an AI system which would likely fall under the Korean AI act and within the scope of Art 3 of the EU AI Act.

Concerning whether the AI can be considered high-risk, the WG1 found that it is quite possible that the XPRS AI Tool to Predict Type 2 Diabetes may be classified as high-impact system under the Korean AI law, should it be fully developed and used.

The WG1 found that *the research project* itself is not regarded as a high-impact AI case. Nevertheless, the developers might want to consider this issue, as it is important because it is reasonable to expect that the AI system, or tool, will ultimately be developed and further tailored to be commercially available as medical or other AI service; or deployed by hospitals for clinical purposes, for instance.

WG1 found that the XPRS system in question (if/when adopted/deployed in the actual medical environment) may fall within one of the sub-provisions for high-impact cases set out in Korean law:

1. Establishment and operation of a system for provision and use of health care services pursuant to Article 3, Paragraph 1 of the Basic Health Care Act[5]

2. Development and use of medical devices according to Article 2, Paragraph 1 of the Medical Devices Act and digital medical devices according to Article 2, Paragraph 2 of the Digital Medical Products Act

3. Other areas that have a significant impact on the safety of human life and body and the protection of basic rights, as determined by Presidential Decree

While not applicable in Korea, WG1 also found that the PRS/XPRS system would fall squarely under Annex III, Section 5 (b) of the EU AI Act high-risk systems if it were integrated into a diagnostic, preventative, or treatment pathway or influences medical recommendations. This seems likely as it is intended for healthcare risk prediction, including early identification and stratification and can influence medical decision-making (e.g., lifestyle interventions, closer monitoring, pharmacotherapy) with processing of sensitive genetic data to evaluate individual health risks.

---

[5] The EU AI Act Art 3 defines AI as "(1) 'AI system' means a machine-based system that is designed to operate with varying levels of autonomy and that may exhibit adaptiveness after deployment, and that, for explicit or implicit objectives, infers, from the input it receives, how to generate outputs such as predictions, content, recommendations, or decisions that can influence physical or virtual environments."

When considered a high-risk AI system under Article 6 of the EU AI Act, the deployer has several obligations and requirements. Among these are the requirement to perform an assessment of the impact on fundamental rights that the use of such a system may produce, as set out in Art 27 of the Act. In Korea, the AI Basic Act defines the class of high-impact AI system (Article 2(4)) that is equivalent to the EU category of high risk. As discussed in the previous section, the business operators providing high-impact systems or incorporating such systems into their products or services are required to ensure AI transparency (Article 31), meet safety and reliability requirements, as well as to conduct impact assessment (Article 35).

**National Guidelines for AI Ethics and a brief comparison with European counterpart**

The National Guidelines for AI Ethics consists of three grand principles and ten requirements designed to be applied across different aspects of personal life, industries, and government activities. During the group discussion within WG1, we exchanged ideas about how to address potential challenges of using our existing analytical framework that is primarily European in its origin in consideration with local contexts. Differences in political, economic, and socio-cultural understanding of the country (or the regional bloc) can affect the scope and contents of ethical principles and their interpretation. With this concern in mind, we continue our discussion with the hope of establishing a reasonable analytical framework suitable for the current project.

Based on a close analysis of the National Guidelines for AI Ethics and its European counterpart, we find that the Korean approach to AI ethics is aligned well with the EU Ethics guidelines for Trustworthy AI. To the point where these two are sufficiently compatible and do not conflict directly, it is safe to conclude the following: even when using the existing framework embedded in prior Z-inspection® practice, this will not lead to inappropriate and unreasonable assessment of what the XPRS case entails from the perspective of applicable law and ethical principles introduced in the domestic setting. While principles such as transparency, accountability, and the protection of privacy are phrased identically across the two documents, the gist of the principles like fairness in the European guidelines are broken down into multiple principles including respect for diversity, solidarity and public good in the Korean document. The overall structure and the core understanding of each guideline seem to be highly compatible. This should not be a surprise; the Ministry (MIST) explicitly noted that when developing Korean guidelines, it sought to consider the global landscape of AI ethics discourse by incorporating EU Ethics guidelines for Trustworthy AI and the OECD AI Principles, both adopted in 2019.

The National Assembly of Korea adopted the AI Basic Act in December 2024.[6] With this passage, Korea became the second country with an AI-specific law that encompassed a comprehensive regulatory scope. Since the government revealed initial drafts of the enforcement decree and guidelines for this Act in September 2025, the contents of many provisions will soon be clarified.[7] There will also be the cases of suspension and grace period for administrative fines. The AI Basic Act has some common aspects shared with the EU AI Act. However, its scope of obligations is far narrower than EU legislation, and enforcement measures are highly limited. Unlike in the EU case, AI

---

[6] The AI Basic Act of Korea (National Assembly of Korea, 2025) [25] consolidates 19 different AI bills introduced to the National Assembly since May 2024, after the 22nd session of the National Assembly began.

[7] See (NIA Korea, 2025) [24] for the draft enforcement decree to the AI Basic Act of Korea.

Basic Act does not have a provision concerning "unacceptable risk", which is a category for the prohibited use and development. Korean law sets out different criteria for categories of AI such as high-impact AI, high-performance AI, and generative AI. Among other things, we find a clear relevance of provisions concerning high-impact AI that is equivalent to the EU category of high risk.

According to Article 2(4) of the AI Basic Act, high-impact AI refers to AI systems that could "significantly affect or pose risks to human life, physical safety and fundamental rights." There are sub-provisions to Article 2(4) setting out the key areas of concern such as energy and water supply, public health/medical, the collection and use of biometric information in the law enforcement context, student assessment by primary education institutions, safety and management of nuclear facilities and more. Our case of the XPRS AI tool may fall within the definition of Article 2(4). More precisely, its sub-provision concerning AI products used for public health and medical treatment is applicable. Existing laws such as the Medical Devices Act and Digital Medical Products Act are also mentioned as a reference point of regulation.

However, the overall applicability is only minimal in the current form and developmental status of the XPRS AI tool. Unless the tool is incorporated into a specific medical device and/or implemented for clinical use, it is unlikely that the mere development of the AI tool itself (as in our case) will be subject to the AI Basic Act. When incorporated and/or implemented in the world of medical practice, for instance, the business operators need to comply with AI transparency (especially prior notification)[8] and safety and reliability requirements[9] (e.g., developing and implementing a risk management plan, developing and implementing a user protection plan, and ensuring human supervision and oversight of such systems) as well as to conduct impact assessment[10] (yet not as strict legal duties in view of Article 35 and without strong enforcement measures attached).

*Other relevant Korean legislations that are not AI-specific*

There are some legislations that are not AI-specific, but still important to examine the potential impact coming from the development and possible application of the XPRS AI tool. The Personal Information Protection Act of Korea is one such example. Our case is concerned with the extensive collection and processing of genetic data involving thousands of people. Genetic information contains inherited characteristics of a natural person which give unique information about physiological features and health conditions of that person. It is thus classified as sensitive information within the meaning of the Personal Information Protection Act. Since this report does not intend to provide our view on legal compliance, we do not plan to engage in further investigation into these laws' applicability in practice. Yet, it should be noted that the use and development of XPRS need to be approached with caution as this AI tool analyzes and makes use of a large amount of genetic information.

---

[8] AI Basic Act of Korea (National Assembly of Korea, 2025) [25], Article 31.

[9] AI Basic Act of Korea (National Assembly of Korea, 2025) [25], Article 34.

[10] AI Basic Act of Korea (National Assembly of Korea, 2025) [25], Article 35.

**The inclusion of a human rights assessment, using the Council of Europe's HUDERIA tool and international human rights standards applicable in Korea.**

WG1 integrated elements of the HUDERIA[11] risk and impact assessment of AI systems in its considerations of Human Rights Law issues related to the AI system.

While HUDERIA is a methodology developed and adopted by the Council of Europe, it is intended to play a role at the intersection of international human rights standards, more broadly, and existing technical frameworks on risk management in the AI context and can therefore be used by a broad set of both public and private actors globally to aid in identifying and addressing risks and impacts to human rights, democracy and the rule of law throughout the lifecycle of AI systems. Specifically, HUDERIA aims to assist States assessing whether they meet their obligations under the Framework Convention on Artificial Intelligence and Human Rights, Democracy and the Rule of Law [27].

A comparison of HUDERIA with the Z-Inspection® found that:

- Both methodologies consider the socio-technical scenarios and complexities across the AI lifecycle. This approach allows for considerations of the interconnected relationship of technology, human choices, and social structures, and looks at how the AI will or might be used in practice.

- Both approaches have trustworthy AI as a stated goal, HUDERIA further makes the standards specific with reference to human rights, while the Z-Inspection® method can be adapted to different standards and normative frameworks, human rights being among these.

- HUDERIA helpfully builds on well-known variables, concepts, and language for the assessment of risks to human rights (scale, scope, probability, and reversibility of potential adverse impacts on human rights). Though HUDERIA aims at mapping potential adverse impacts on human rights, the framing can also support an approach looking at both a positive and adverse impact on human rights, something Z-Inspection® found was useful in its pilot use case on fundamental rights [28].

- The four steps of HUDERIA 1. the Context-Based Risk Analysis (COBRA), 2. the Stakeholder Engagement Process (SEP), 3. the Risk and Impact Assessment (RIA) and 4. the Mitigation Plan (MP)) have similarities, overlaps and complements the Z-Inspection® process.

- The COBRA overlaps with the Z-Inspection® socio-technical scenario approach and introduces a mapping of potential adverse impacts on human rights related to the system's intended *application* context, *design and development* context and d*eployment* context. These loosely correspond to the Z-Inspections® working groups: technical (design and development), legal/ethical (application) and domain (deployment), in this case healthcare.

---

[11] The Council of Europe risk and impact assessment of AI systems from the point of view of human rights, democracy and the rule of law ("the HUDERIA") provides a structured approach to risk and impact assessment for AI systems specifically tailored to the protection and promotion of human rights, democracy and the rule of law. See (HUDERIA Methodology, 2024) [26].

- The SEP step reflects the interdisciplinary nature of the Z-Inspection® and the identification of stakeholders as part of the socio-technical scenario and introduces a more rigorous approach to stakeholder analysis.

- The two final steps RIA and MP have similarities with the open language identification of issues in the Z-Inspection® and the mapping to requirements and suggestions for how to mitigate or solve tensions.

In this use case, four steps were applied by the WG1 to identify relevant human rights issues, using also the HUDERIA:

1. Open language conversations in the group on ethical risks and concerns

2. AI generated COBRA (Context-based Risk Analysis, first step in the HUDERIA Methodology) assessment of the AI for HUDERIA

3. Small group review of the AI generated assessment, the HUDERIA framework and the open language issues to identify most relevant points.

4. Discussion and validation of the identified points in the full WG1.

## Human Rights Law findings

*Positive contributions*

WG1 first stressed the importance of considering what potential *positive contribution* the XPRS AI system could have towards the achievement of human rights and not just any potential threats or human rights violations.

Two human rights were identified: t*he human right to health*,[12] and a *human right to benefit without discrimination from the fruits of scientific discovery*.[13] It was noted that AI technologies have the potential to dramatically reduce inequalities across the wealth gradient, making better diagnoses possible in areas that might otherwise lack the resources.

*Findings about the system in HUDERIA categories for context-based risk analysis*

As to the risk factors relating to the system's application context (so-called COBRA A), the following findings have been made:

- **Public administration**: Healthcare and research

---

[12] The 1948 Universal Declaration of Human Rights also mentioned health as part of the right to an adequate standard of living (art. 25). The right to health was again recognized as a human right in the 1966 International Covenant on Economic, Social and Cultural Rights.

[13] See the right to access to, to participate in and to enjoy science and its benefits (OHCHR, 2023) [29], as well as the international human rights legal framework which contains international instruments to combat specific forms of discrimination, also in the area of health.

- **Essential services offered by the private sector**: Biomedical applications and healthcare, potentially health insurance.

- **Sector**: Healthcare is a safety-critical, high-risk domain. The system, if introduced for clinical use, would inform medical decision-making, implicating physical and mental integrity. It may further inform research and public health.

- **Function**: The AI system performs a high-impact screening function. While not directly diagnostic, depending on the modalities of its deployment it has the potential for substantially influencing clinical interventions. The polygenic risk score (PRS) system will likely play a significant role in a high-impact and safety-critical context, particularly within the healthcare domain. Its outputs may influence important medical, psychological, or preventative decisions affecting individuals' access to care, early interventions, or lifestyle adjustments.

- **Regulatory context**: Regulated under South Korea's Medical Devices Act if used in clinical settings; also subject to PIPA for genomic data. EU deployment would trigger GDPR and possibly AI Act high-risk classification.

- **Legal basis**: Must ensure clear legal basis for data processing (consent and purpose limitation) and medical deployment (if considered a software as medical device, SaMD).

- **Overlapping IAs**: Likely requires a Data Protection Impact Assessment (DPIA) under GDPR/Convention 108+. May also require ethics board approval.

- **Repurposing risk**: High. Output could be misused for insurance underwriting or employment screening. Needs strict purpose limitation and data sharing controls.

- **Human Rights Due Diligence**: Requires due diligence on bias, privacy, consent, and health equity. Stakeholder engagement should include patient advocates and minority populations.

- **Scale of deployment**: National or multinational scale. Potential to affect millions of individuals, even pre-symptomatically.

- **Number of people affected**: Depends on screening population but can reach millions. Disproportionate effects are likely among underrepresented groups.

- **Timescale of impact**: Long-term: Impacts may last decades; risk profiles can follow individuals throughout life.

- **Historical discrimination**: Genomic datasets underrepresent non-European and rural East Asian populations. Structural health bias risk is high.

- **Vulnerable groups**: Yes: groups of individuals - non-Korean and non-Japanese population - potentially underrepresented because of the way in which the system has been trained and built, other minorities, low-income groups, elderly, and people with limited access to genomic testing may be disadvantaged.

- **Children**: If used for family-based risk prediction, children may be indirectly affected, requiring special considerations and measures.

### Identified Human Rights Issues and Tensions (WG1)

For the identification and mapping of potentially relevant human rights issues a ChatGPT supported list of issues was identified based on the yet unpublished HUDERIA COBRA Resources. These issues were then reviewed, discussed and curated by WG1, in smaller groups. While the AI generated input was rich in content, significant revisions were required both to correct errors and tailor input to the case. As the input was AI generated the discussions and the review were complicated, as none of the WG1 participants could explain the input or had ownership over it.

Four of the HUDERIA-generated "areas of potential concern" were identified by WG1.

1. *Physical and mental integrity and human dignity:*

The physical and mental integrity and human dignity are ethical cornerstones of the international order and human rights provisions provide a number of protections regarding the physical and mental integrity of individuals, reinforcing the importance of personal and individual autonomy and self-ownership. Many of the legal provisions contribute to protecting physical and mental integrity and human dignity, including the right to life, to health, to non-discrimination and to be free from torture, inhumane and degrading treatment.[14] Human rights protections may extend to various other situations, where physical and mental integrity of individuals may be at stake.

Nevertheless, references to human dignity, such as in the Universal Declaration of Human Rights, are not directly binding [31]. In the European context, the EU Charter of Fundamental Rights has a specific provision regarding human dignity in Article 3. For the purpose of this report, WG1 considered the right to physical and mental integrity and human dignity in a broader sense.

A polygenic risk score (PRS) system may play a significant role in a high-impact and safety-critical context, and its outputs may influence important medical, psychological, or preventative decisions affecting individuals' access to care, early interventions, or lifestyle adjustments. As such, PRS may trigger clinical monitoring or intervention pathways, including potentially invasive procedures or pharmacological regimens, all of which could affect mental health by shaping individual risk perceptions or influencing medical decisions. For example, a patient may suffer significant psychological, financial, and social impacts based on information generated by the PRS, and further suffer significant trauma, especially if that information later proves to be either unsubstantiated or more alarming than scientifically warranted.

It follows that XPRS, which sets out to explain the PRS more readily, will have similar risks. In its function, therefore both PRS and XPRS may directly or indirectly implicate physical and mental integrity and human dignity.

With this in mind, WG1 recommends due consideration to:

1) the high demands placed on ensuring accuracy in prediction

2) measures needed to ensure autonomy for clinicians and patients

3) safeguarding the right to choose its use, including non-use

---

[14] Article 7 (Prohibition of torture or cruel, inhuman or degrading treatment) of the ICCPR (UN General Assembly, 1966) [30].

4)    informed decision-making

This should be taken into consideration if or when used, so the risk can be mitigated both through regulations, training, and protocols for use.

[Due to its existential high impact (life-changing consequences), the stakeholders planning to use PRS and XPRS should commit not only to (1) maximizing the precision of predictions, (2) protect the autonomy of the persons concerned and (3) respect their freedom of choice, but also (4) ensure that all persons (doctors and especially patients) have a real opportunity to make an informed decision. In other words, the autonomy and freedom of choice of the individuals concerned must not only be formally guaranteed but should also be proactively supported by providing information about possible (positive and negative, health-related and social) consequences. [This aspect of proactive information-policy to support an informed decision (especially) by the patient is basically inherent in the concepts of 'autonomy' and 'freedom of choice', but can be ignored under conditions of time constraints and structural asymmetries (see the epistemic asymmetry between doctor and patient), so that it is important to explicitly remind of this fourth point here. The implementation of a proactive information policy is the responsibility of the respective concrete institutional structures that are responsible for the implementation of AI systems.]

2.    *Privacy and data protection*

Privacy and data protection rights ensure limits to outside influence over private affairs and that people can keep their personal data, behaviors, and decisions from being disclosed or monitored without their consent, safeguarding personal autonomy and dignity. Specifically in the public sector context, such as health, privacy and data protection laws protect individuals' personal information by ensuring that public authorities or entities acting on their behalf only process personal information that they are authorized to. Privacy protections also allow individuals to maintain control over how they grow, define and present their personal identity. These protections protect against unauthorized use or misrepresentation of an individual's personal attributes, such as name, likeness, or other distinguishing characteristics.

The right to privacy is established in several human rights instruments, including Article 12 of the [Universal Declaration of Human Rights] and Article 17 of the [International Covenant on Civil and Political Rights].

In this case, the WG1 considered that PRS relies on large databases containing DNA on a given population, which allowed for PRS research and the development of the XPRS. It was further noted that DNA testing for an individual would be required for the PRS and the XPRS to be used clinically in the treatment of individuals.

Both the large database use and the individual DNA testing raise some privacy and data protection related issues and as the use of genetic, biometric, or health data invokes strict obligations, such as explicit informed consent, purpose limitation and data minimization, and data security and access controls. The XPRS system presents multiple incentives for malicious exploitation, particularly due to the sensitivity, value, and potential influence of the data and outputs it generates. In response to these incentives, we recommend that researchers embrace and adopt best practices regarding data security and encryption for any sensitive data collected.

3. *Equality and non-discrimination*

The healthcare and genomics sectors, along with the data used to train AI-enhanced polygenic risk score (PRS) systems, may contain documented patterns of bias, inequality, and under representation, particularly with respect to: 1) ethnic and racial minorities, 2) women and non-binary individuals, 3) low-income and marginalized populations, and 4) persons with disabilities or rare diseases.

The replication of existing patterns of discrimination and inequality is a real concern unless explicitly mitigated. Key risks include:

[Short-Term Impacts:] Lower predictive validity for non-European/*non-Korean and non-Japanese* populations can lead to misdiagnosis or lack of clinical utility; psychological harm or false reassurance based on unreliable predictions; and insurance or employment decisions based on flawed or poorly understood risk scores.

[Medium-Term Impacts:] Algorithmic feedback loops, where biased outputs are incorporated into future care recommendations; erosion of trust in genetic technologies by marginalized groups; and inequitable access to AI-enhanced preventative care.

[Long-Term Impacts:] Entrenchment of biological essentialism, reinforcing racial or gender stereotypes; genomic discrimination in insurance, education, and employment sectors; and system-wide inequality amplification in population health outcomes.

More would be required to ensure that potential risks of bias and indirect discrimination in the context of model development and model implementation are fully mitigated. This could be through continued validation and evaluation of the trained AI model. Possibly also investment, based on proof of concept in the Korean context, in other medically and scientifically underserved contexts.

4. *Children*

Both general human rights and children-specific protections exist in respect of this group. They are aimed at ensuring that children can grow, learn, play, develop and flourish with dignity and that children's specific vulnerabilities and needs are properly considered. In all actions concerning children, their best interests must be a primary consideration.

Using polygenic risk score (PRS) systems to predict diabetes risk in children raises serious concerns around explainability and age-appropriateness. These models combine thousands of genetic variants in opaque ways, making their outputs difficult to interpret even for clinicians. For families and young patients, results can easily be misunderstood as deterministic, creating unnecessary anxiety or a false sense of certainty. Children are also in constant biological and psychological development, which means that risk estimates at a young age may not carry the same predictive weight over time, and the very act of disclosing them can affect a child's well-being and self-perception.

For these reasons, PRS outputs for children require careful professional interpretation and must never be delivered without clinical framing. The scores should be contextualized with lifestyle, environment, and family history, and explained in language appropriate to the child's age and understanding. Without these safeguards, the system risks undermining trust, stigmatizing children, or driving inappropriate interventions, rather than contributing to their health and long-term interests.

**Identified Ethical Issues and Tensions (WG1)**

*Ethical framing and issues*

At the core of Z-inspection® is considering the long-term societal implications and fairness, sustainability and societal well-being and impact of an AI system being used, by looking at socio-technical scenarios. Both the PRS use for screening and individual prediction and the XPRS for explanation by clinicians were considered.

Ethical issues were considered based both on the EU Ethics Guidelines for Trustworthy Artificial Intelligence and the South Korean ethical guidelines.[15] A comparison showed that both guidelines have "trustworthiness" as the central concept. This is in the Korean context set out in 10 principles[16], and for the European in seven requirements.[17] While human rights protection is directly included as a principle of ethical AI in Korea, it is implicit for the European guidelines. When comparing the EU Ethics Guidelines for Trustworthy Artificial Intelligence with the National Guidelines for AI ethics in Korea, WG1 found that there were no real conflicts between the two and the EU Guidelines were used to align with the language in other Z-Inspection use-cases.

**Ethical issues identified and Mapping of the ethical issues to the Pillars, Requirements, and Sub-requirements**

The group identified a number of ethical issues in the context of the XPRS use case. Some of the ethical issues play an important role both during the research phase and when used in the clinic, others play out only in case the tool is used in clinical practice.

---

[15] The National Guidelines for AI Ethics (Ministry of Science and ICT, 2020) [32], prepared by the Ministry of Science and ICT (MIST) and adopted in December 2020, were designed to introduce general principles applicable across different fields. They consist of three principles (must be considered in the process of developing and using AI for humanity) and ten requirements (to follow those three principles, key requirements that must be met throughout the AI system's lifecycle). The Guidelines explicitly noted that the Ministry sought to incorporate EU Ethics guidelines for Trustworthy AI and the OECD AI Principles (both adopted in 2019) in consideration with the global landscape of AI ethics discourse at that time.

[16] The 10 principles (our translation) 1) human rights protection, 2) privacy, 3) respect for diversity, 4) prohibition of infringement/ "do no harm", 5) respect and promotion of public values/common good, 6) solidarity, 7) data management, 8) accountability, 9) safety and 10) transparency.

[17] The seven requirements of the Ethics Guidelines for Trustworthy AI (AI HLEG, 2019) [20]: 1) Human agency and oversight (Including fundamental rights, human agency and human oversight), 2) Technical robustness and safety (Including resilience to attack and security, fall back plan and general safety, accuracy, reliability and reproducibility), 3) Privacy and data governance (Including respect for privacy, quality and integrity of data, and access to data), 4) Transparency (Including traceability, explainability and communication), 5) Diversity, non-discrimination and fairness (Including the avoidance of unfair bias, accessibility and universal design, and stakeholder participation), 6) Societal and environmental wellbeing (Including sustainability and environmental friendliness, social impact, society and democracy), 7) Accountability (Including auditability, minimization and reporting of negative impact, trade-offs and redress).

In a second step, the ethical issues identified were then mapped against the four ethical pillars and seven requirements and sub-requirements outlined in the Ethics Guidelines for Trustworthy AI [20]. The four ethical pillars are Respect for Human Autonomy, Prevention of Harm, Fairness, and Explicability.

For most of the identified ethical issues, several mapping options exist, some of which are depicted below.

- **Data used**: It is understood that the PRS are done based on large scale Japanese and Korean data sets. What limitations might this raise when used on other groups?

| Ethical Pillars | Requirements | Sub-requirements |
|---|---|---|
| Prevention of Harm | Diversity, Non-discrimination, and Fairness | Avoidance of Unfair Bias |
| Fairness | Technical Robustness and Safety | Accuracy, Reliability |

- **Role of genetic information:** As the PRS is based on genes only, how does it include and weigh risks from other factors in the score? Are there debates and uncertainty about the underlying science on the basis of which the PRS are developed? (transparency, privacy, robust, reliable, discrimination, explainability)

| Ethical Pillars | Requirements | Sub-requirements |
|---|---|---|
| Prevention of Harm | Transparency | Communication |
| Prevention of Harm | Technical Robustness and Safety | General Safety; Accuracy; Reliability |

- **Role of Explainability**: What is the role of explainability in the approach? What is the quality of the explanation provided? Does the addition of explainability reduce accuracy? Do doctors and patients understand the explanation?

| Ethical Pillars | Requirements | Sub-requirements |
|---|---|---|
| Explicability | Accountability | Risk Management |
| Explicability | Technical Robustness and Safety | Accuracy; Reliability |

- **Client relation, behavior and vulnerability**: How to deal with the PRS being a risk score (prediction), not a diagnosis? Other indicators and variables like weight etc. could be relevant. How might risk change over time - how does it develop? (reliability, transparency, autonomy, do no harm, explainability)

| Ethical Pillars | Requirements | Sub-requirements |
|---|---|---|
| Respect for Human Autonomy | Technical Robustness and Safety | Reliability; Accuracy |
| Explicability | Human Agency and Oversight | Human Agency and Autonomy |
| Prevention of Harm | Transparency | Explainability; Communication |

- **Informed consent:** How was consent ensured for the collection and use of genetic data, for what, and for how long? Do you want to give DNA data? (privacy, autonomy, equality, fairness)

| Ethical Pillars | Requirements | Sub-requirements |
|---|---|---|
| Respect for Human Autonomy | Privacy and Data Governance | Privacy |
| Respect for Human Autonomy | Diversity, Non-Discrimination and Fairness | Stakeholder Participation |

- **Right not to know**: Do individuals have the right to say no to DNA mapping, and is this option offered in a transparent and adequate way? How would this be done with individual genetic testing, how with genetic screening? How can individuals' right not to know be granted? What are options for individuals not to have the test taken?

| Ethical Pillars | Requirements | Sub-requirements |
|---|---|---|
| Respect for Human Autonomy | Privacy and Data Governance | Privacy |
| Respect for Human Autonomy | Transparency | Explainability; Communication |

- **Amount of information individuals receive**: Will individuals learn about their susceptibility to diabetes type II or will they learn about their entire genome? Can they decide for themselves how much information to receive?

| Ethical Pillars | Requirements | Sub-requirements |
|---|---|---|
| Respect for Human Autonomy | Privacy and Data Governance | Privacy; Data Governance |
| Respect for Human Autonomy | Transparency | Traceability; Explainability; Communication |

- **Data use and privacy - do no harm:** Can researchers use the genetic data for other research studies? Can they sell it? Who can get access to the DNA data in general and specifically? Can insurance companies get access? How and how long is the data stored?

| Ethical Pillars | Requirements | Sub-requirements |
|---|---|---|
| Respect for Human Autonomy | Privacy and Data Governance | Privacy; Data Governance |
| Prevention of Harm | Privacy and Data Governance | Privacy; Data Governance |

- **When and how to tell a patient - autonomy**? Risk is a continuous variable - then when do you tell it?

| Ethical Pillars | Requirements | Sub-requirements |
|---|---|---|
| Respect for Human Autonomy | Transparency | Traceability; Explainability; Communication |

- **Social aspects**: Need to consider the existing interdisciplinary debate on ethical and social aspects of genetic testing and genetic screening. This general aspect was not mapped, as it covers a broad spectrum of ethical and societal aspects that go beyond the Ethics Guidelines for Trustworthy AI.

## Lessons learned WG 1 Human Rights and Ethics: Using HUDERIA and Z-inspection®

In the context of the use case, Z-inspection® and HUDERIA complemented each other.

The Z-inspection® methodology provided the background for an open and flexible discussion of practical, ethical, and human rights aspects related to the PRS use case. Through a series of working group discussions, WG1 team members contributed their thoughts and the WG jointly developed a list of open language practical, ethical and human rights issues and tensions to be considered in the context of the PRS use case and similar use contexts. Throughout the project, the quality and depth of the arguments increased. The project meetings allowed for an interdisciplinary exchange that increasingly sharpened the WG perspective and helped the WG1 to come up with a number of ethical issues and tensions.

In contrast, HUDERIA is a structured, legally anchored analytical framework for identifying and assessing human rights-related risk factors. In this pilot, the COBRA component (based on COBRA resources adopted by the Council of Europe on 12 February 2026) was done using AI-support to generate structured sample responses across three analytical dimensions: (1) risks arising in the AI system's use context, (2) risks arising in its design and development context, and (3) risks arising in its deployment context. The output was detailed and presented in a systematic format (short statements and bullet points), explicitly oriented towards human rights standards and relevant legal frameworks. In some instances, the responses referred to specific national or international legal instruments; in others, they articulated more general principles derived from applicable human rights and data protection norms.

The AI-generated responses did not constitute a final assessment. Rather, it functioned as a structured risk-mapping tool that required careful review and contextual validation by the working group. Certain elements were framed at a high level of generality, while others required clarification as to their practical relevance in the specific PRS scenario. However, the structured mapping also surfaced considerations that had not emerged in earlier discussions and provided a coherent architecture for organizing the group's human rights analysis.

As such, the AI-supported component was used as an experimental tool to engage with the COBRA framework efficiently and to obtain an initial structured mapping of potential human rights risks, particularly considering limited time and expert resources. It allowed the working group to quickly familiarize themselves with the range of questions embedded in the methodology. At the same time, the AI responses revealed important limitations, including elements of generality and uneven contextual precision. Consequently, substantial expert review and refinement were necessary to validate, prioritize, and adapt the findings to the specific PRS use case.

Overall, the exercise demonstrated that HUDERIA and Z-Inspection® operate at different but complementary levels. Z-Inspection® enabled in-depth, deliberative exploration of ethical tensions and contextual nuances, while HUDERIA contributed a systematic, legally grounded structure for identifying and categorizing potential human rights risks. A thorough and balanced human rights assessment required the combination of both elements: structured analytical scaffolding and interdisciplinary expert judgment. However, in itself, the AI-generated sample answers did not offer a thorough and pointed assessment of the use case. For a detailed analysis that provides a balanced assessment of the use case from a human rights perspective, considerable additional human input is required. Given the Z-inspection® based WG discussions on ethical and human rights aspects of the use case, the WG was able to build on the AI-generated output, identify the most relevant points, and validate them from the perspective of WG1.

## IV. RELATED WORK

In this section, we cover the Literature/Background Study/Evidence Base and we look for similarities.

### Literature/Background study/Evidence Base

Polygenic risk scores (PRS) have emerged as a powerful tool for quantifying genetic susceptibility to complex diseases by aggregating the effects of numerous common genetic variants across the genome [33, 34]. With the rapid expansion of genome-wide association studies (GWAS) and biobank-scale datasets, PRS have demonstrated increasing potential for disease risk prediction, population stratification, and personalized prevention strategies [35, 36]. Consequently, PRS are being actively explored as a core component of precision medicine initiatives, including clinical risk stratification, early screening, and tailored interventions [37, 38].

Despite these advances, the clinical translation of PRS remains controversial. While PRS can enhance predictive performance beyond traditional risk factors, concerns persist regarding their stability, generalizability across ancestries, ethical implications, and real-world clinical utility [43, 35, 33, 34, 39]. In particular, discrepancies in risk classification across different PRS constructions, limited transferability to non-European populations, and unresolved privacy and regulatory challenges pose significant barriers to widespread adoption [44, 43, 35, 33, 40].

### Benefits of Genetic Data and PRS

#### 1. Enhanced Predictive Power

PRS improves disease risk prediction by aggregating the cumulative effects of thousands to millions of genetic variants, capturing polygenic architectures that are not identifiable through single-variant analyses [33, 34]. Empirical studies demonstrate that integrative approaches such as PRS mix and PRS mix+ can significantly enhance predictive accuracy, with reported improvements of approximately 1.2–1.7-fold in European and South Asian populations compared to conventional PRS methods [41]. These gains highlight the value of leveraging heterogeneous genetic architectures and optimized weighting schemes for risk prediction.

Furthermore, integrating PRS with clinical and phenotypic data within secure data-sharing frameworks has been shown to improve cohort-level analyses and enable more refined risk modeling for precision medicine applications [37, 38]. Such integrative models outperform genetic-only or clinical-only approaches by capturing complementary sources of disease risk [33, 36].

#### 2. Personalized Risk Assessment

PRS enables individualized genetic risk stratification by placing individuals along a continuous risk distribution rather than binary disease categories. This stratification supports targeted prevention strategies, including intensified screening, early intervention, and lifestyle modification for high-risk individuals [33, 35]. In some diseases, individuals in the highest PRS percentiles exhibit risk levels comparable to those conferred by rare monogenic variants [42].

### 3. Cost-Effective Screening

Compared to sequencing-based diagnostics or rare-variant testing, PRS offers a scalable and cost-effective approach for population-level risk identification once genotyping infrastructure is established [33, 38]. The marginal cost of PRS computation is low, enabling large-scale screening programs and public health applications [36].

### 4. Cross-Ancestry Utility

Recent methodological advances, including multi-ancestry GWAS integration and approaches such as PRS mix+, have improved PRS transferability across diverse populations [41, 35]. These developments partially mitigate Eurocentric biases and hold promise for reducing health disparities in genetic risk prediction [35, 33].

## Disadvantages of Genetic Data and PRS

### 1. Instability in Risk Classification

High-risk classification based on PRS is sensitive to methodological choices, including GWAS source, variant selection, and modeling framework [43, 33]. Recent evidence shows that individuals classified as high risk by one PRS may not be similarly classified by alternative scores, resulting in substantial instability at the individual level. This inconsistency undermines confidence in PRS-guided clinical decision-making [43, 34].

### 2. Limited Generalizability

Most PRS are trained predominantly on European-ancestry GWAS, leading to reduced predictive accuracy in non-European populations [35, 33]. Differences in linkage disequilibrium structure, allele frequencies, and genetic architecture further exacerbate this limitation [35]. Moreover, PRS performance varies by trait, reflecting differences in heritability and polygenicity [35, 34].

### 3. Privacy and Ethical Concerns

Genetic data remain inherently identifiable even after deidentification processes, raising concerns about privacy breaches and unauthorized re-identification [44, 45, 39]. These risks are compounded by inconsistent regulatory frameworks governing genetic data use, particularly in direct-to-consumer (B2C) genetic testing and cross-border data sharing [37, 44, 40, 46].

### 4. Methodological Limitations

Most PRS models do not explicitly incorporate environmental exposures, lifestyle factors, or gene–environment interactions, limiting their explanatory power for complex diseases [33, 34]. Additionally, advanced PRS methods often require large training datasets and substantial computational resources, posing challenges for reproducibility and clinical implementation [34, 35].

### 5. Clinical Utility Gaps

Although PRS can achieve statistically significant associations with disease outcomes, absolute prediction accuracies often fall short of thresholds required for clinical actionability [34, 36]. Furthermore, the absence of standardized metrics for individual-level uncertainty or error estimation limits clinician trust and interpretability [34, 42].

## Comparisons to Other Diseases

### Similarities: Where PRS + Explainability Is Applicable

The integration of polygenic risk scores (PRS) with explainability frameworks (e.g., XPRS) is particularly well suited for complex diseases characterized by high polygenicity, where disease risk arises from the cumulative effects of many common variants with modest individual effect sizes. In such settings, explainable PRS approaches enable not only robust risk stratification but also biologically interpretable decomposition of genetic risk into pathways or gene sets, thereby facilitating clinically meaningful insights beyond aggregate risk estimates.

### Cardiovascular Disease (CVD)

Cardiovascular disease, particularly coronary artery disease (CAD), represents one of the most established use cases for PRS. Large-scale population studies have demonstrated that PRS can identify individuals at substantially elevated CAD risk, often comparable to that conferred by rare monogenic mutations, especially among younger individuals without traditional clinical risk factors [59, 7]. In this context, explainability tools can further enhance clinical utility by decomposing PRS into interpretable biological components, such as lipid metabolism, inflammation, and vascular remodeling pathways. Such pathway-level interpretation supports individualized understanding of genetic risk and may inform targeted prevention strategies [7].

### Hypertension

Hypertension exhibits a strongly polygenic genetic architecture, similar to other cardiometabolic traits, with risk distributed across hundreds of loci with small effect sizes. Large-scale genome-wide association studies involving over one million individuals have identified more than 500 loci associated with blood pressure traits, underscoring the suitability of PRS-based modeling for early risk stratification [47]. Explainable PRS frameworks are particularly valuable in hypertension, as they can help distinguish whether renal, neurohormonal, or vascular mechanisms predominantly contribute to an individual's genetic risk. This mechanistic stratification aligns with the known clinical heterogeneity of hypertension and supports precision prevention and management strategies [47].

### Differences: Where PRS + Explainability Has Limited Value

By contrast, polygenic risk scores (PRS) combined with explainability frameworks such as XPRS have limited incremental value in disease contexts where genetic risk is dominated by rare, high-penetrance variants or where the contribution of common polygenic variation is relatively small. In such scenarios, disease risk is largely discrete rather than continuous, and clinical decision-making relies primarily on rare-variant detection and variant-level interpretation rather than on polygenic aggregation or score decomposition.

### Breast and Ovarian Cancer (BRCA1/2-driven)

Hereditary breast and ovarian cancer (HBOC) represents a disease context in which genetic risk is predominantly driven by rare pathogenic variants with high penetrance, most notably in BRCA1 and BRCA2 [48]. Large cohort studies have demonstrated that carriers of pathogenic BRCA1/2 variants have substantially elevated lifetime risks of breast and ovarian cancer, establishing these variants as the principal determinants of clinical risk stratification and management [48].

Although polygenic risk scores have been shown to modify cancer risk among BRCA1/2 mutation carriers, the magnitude of this modification is modest relative to the dominant effect of the causal variant itself [63, 60]. As a result, PRS functions primarily as a secondary risk modifier rather than a primary predictive tool in HBOC. In this setting, decomposing PRS via explainability frameworks provides limited additional clinical value, as disease etiology is largely governed by single-gene effects rather than distributed polygenic architecture [63, 60].

**Monogenic Disorders (e.g., Familial Hypercholesterolemia, Cystic Fibrosis)**

For classical monogenic disorders, PRS is not the primary diagnostic or predictive tool because disease susceptibility is overwhelmingly driven by pathogenic variants with large effect sizes in specific genes. Familial hypercholesterolemia (FH), for example, is a hereditary lipid disorder characterized by markedly elevated low-density lipoprotein cholesterol and increased risk of premature coronary artery disease, where clinical management is guided by variant identification, cascade screening, and intensive lipid-lowering therapy rather than by polygenic modeling [49].

Similarly, cystic fibrosis and CFTR-related disorders depend on variant-level assessment, and peer-reviewed technical standards emphasize rare-variant detection and interpretation as the basis for diagnosis and clinical decision-making [50]. In these disease contexts, XPRS approaches offer minimal added value, as decomposing a polygenic score does not substitute for identifying the causal high-penetrance variant or following established variant-driven management pathways [49, 50].

## V. CONCLUSION AND LESSONS LEARNED

This case study assessed the XPRS for T2DM applying the Z-inspection® for Trustworthy AI, a methodology for co-design. As genomic data has been gathered for specific populations, this study focuses XPRS trained and validated on large East Asian cohorts in Korea and Japan, which have collected genomic data in national databases. The lessons drawn are based on the findings from three working groups focusing on technical, medical, and human rights and ethical aspects. The lessons connecting the findings across the working groups consider similar aspects of the XPRS system in identifying the potential risks to trustworthiness of the AI system. First, the framework adopted used those similar to other studies looking at trustworthy AI of systems. Previous studies examined factors that affected trust in AI medicine involving concerns of discrimination, responsibility, and liability [51]. These factors are context dependent, which are not within the scope of the case of the study. These factors are nonetheless important considerations for policy.

Past studies using data from the EU have found that physicians trust AI even when they lack experience and/or training [52]. Medicine, however, is a field that people are most likely to distrust the use of AI [52]. Another study in Japan found that physicians were concerned about adopting AI technology because of issues surrounding accountability [53]. A survey of Korean doctors found that most would use AI for diagnosis but would need human input, especially because AI would oversimplify the information used [54]. These previous findings support those of this study, which suggest that accuracy and accountability are the main points that need to be addressed within the design process for trustworthy AI.

The implications of this study involve what mechanisms should be considered to help ensure that the AI applications are trustworthy and how they might be designed. In the case of XPRS, the medical and technical implications involving utility of the technology suggest that it is still in the developmental stage and not yet ready for clinical use considering trustworthy AI, particularly given the known pitfalls and challenges in genetic test interpretation [57]. The main obstacles from both perspectives center on accuracy and accountability, which require explainability and interpretability. The diagnostic function of XPRS relies on genetic markers, but leaves out many of the traditional factors, such as lifestyle behavior and developmental changes.

Since AI relies on an ecosystem of data collection through to its application, including the integration of diverse genetic databases [62], the scope of trustworthy AI is wide. When considering the diversity of stakeholders, many of them have different perspectives depending on the point of development, which requires consideration of human rights and ethics at each point applying concepts such as ethical layering to address the multifaceted implications of PRS [64]. For instance, while XPRS does not involve data collection directly, it relies on data that has been previously collected third parties. The historical challenge of data withholding in genetics research [58] further complicates the transparency and availability of such data. Additionally, the availability of information generated by XPRS may also prove a liability that requires proper management. From these perspectives, accountability falls well outside the clinician-patient relationship. While WG1 explored some policies in Korea, the EU, and other jurisdictions, these were not comprehensive but nonetheless reflect the breadth of interests that are involved.

The concerns raised regarding the accountability and liability of AI usage reflects a principal agent problem [61]. When disparate business interests are responsible for different elements of the AI system lifecycle that includes a variety of stakeholders, the responsibility and thus liability can become difficult to track without AI governance that includes appropriate traceability measures [55]. Data collection may be performed by a different entity than the XPRS provider, which can be different from the one providing healthcare service or other service derived from the scoring data. This study identified the need to establish thresholds for accuracy and reliability without explicitly defining what those thresholds might be. The relationship network goes well beyond the patient-clinician, which define the different points that require checks on human rights and ethics. Individuals typically expect public policy over industry to ensure trust in AI [52], suggesting a role for government intervention. The complexity suggests that institutions, possibly managed by an agency, are needed to incorporate these perspectives and enable monitoring and even enforcement capabilities [56].

From the researchers' perspective, a central lesson from evaluating the XPRS tool for Type 2 Diabetes is that "trustworthy AI" cannot be equated with a single measure of performance. A model may demonstrate strong discrimination yet remain untrustworthy if its intended use is unclear, its limitations are insufficiently specified, or its outputs are prone to misinterpretation in real-world workflows. We also learned that interpretability is inherently user- and context-dependent. Explanations meaningful to statistical genetics researchers may not translate to clinicians or patients. Accordingly, interpretability should be evaluated relative to explicitly defined target users and tasks, using metrics that assess whether explanations genuinely improve understanding and decision quality.

This work provides a practical framework for future genome–AI projects by clarifying the components required for trustworthiness beyond model training and post hoc explanation.

Specifically, it highlights the importance of (i) defining intended use and target users at the outset, (ii) specifying boundary conditions and "do-not-use" scenarios, (iii) designing evidence-oriented evaluation plans, and (iv) tailoring explanation layers to stakeholder needs. Articulating these elements early in development can reduce overclaiming, prevent late-stage redesign, and better align technical implementation with clinical and ethical standards.

As a next step, informed by reviewer feedback, we plan to conduct a structured user study to empirically evaluate whether the proposed explainability framework improves user understanding, confidence calibration, and decision quality. The results will guide iterative refinement of explanation content and interface design and help formalize which explanatory components are necessary and sufficient for the defined use case. Through this process, interpretability will be operationalized as an empirically validated system property, thereby strengthening the overall case for trustworthiness.

**DISCLAIMER**

The views expressed in this article are solely those of the authors and do not necessarily represent those of their affiliated organizations.

## AUTHOR CONTRIBUTION

WG1 was composed of: Elisabeth Hildt, Emilie Wiinblad Mathez, Heejin Kim, Ralf Beuthan, Stephan Sonnenberg, Vadim Pak, and Sira Maliphol.

WG2 was composed of: Megan Coffee, Na Yeon Kim, Pedro Kringen, Haekyung Lee, and Seunggeun Lee.

WG3 was composed by: Jesmin Jahan Tithi, Magnus Westerlund.

Project Lead and Coordination: Roberto V. Zicari.

## GRANTS

Seunggeun Lee and Na Yeon Kim were supported by the "Brain Pool Plus (Brain Pool+) Program through the National Research Foundation of Korea (NRF), funded by the Ministry of Science and ICT (2020H1D3A2A03100666), and the NRF grant funded by the Korea government (MSIT) (No. RS-2023-00222663)

Heejin Kim was supported by the NRF grant funded by the Korea government (MSIT) (No. RS-2023-00222663).

Sira Maliphol was supported by the New Faculty Startup Fund from Seoul National University.

Roberto V. Zicari was supported by the BrainPool Program through the National Research Foundation of Korea (NRF) funded by the Ministry of Science and ICT (grant number: 2022H1D3A2A01082266, Research Title: Assessing Trustworthy AI).

# REFERENCES

[1] (Sun et al., 2022) Sun H, Saeedi P, Karuranga S, et al., IDF Diabetes Atlas: Global, regional and country-level diabetes prevalence estimates for 2021 and projections for 2045. Diabetes Res Clin Pract. 2022;183:109119. doi:10.1016/j.diabres.2021.109119.

[2] (Wilson et al., 2007) Wilson PWF, Meigs JB, Sullivan L, Fox CS, Nathan DM, D'Agostino RB. Prediction of incident diabetes mellitus in middle-aged adults: the Framingham Offspring Study. Arch Intern Med. 2007;167(10):1068-1074. doi:10.1001/archinte.167.10.1068.

[3] (Hippisley-Cox and Coupland, 2017) Hippisley-Cox J, Coupland C. Development and validation of QDiabetes-2018 risk prediction algorithm to estimate future risk of type 2 diabetes: cohort study. BMJ. 2017;359:j5019. doi:10.1136/bmj.j5019.

[4] (Lee et al., 2025) Lee H, et al., Prediction model for type 2 diabetes mellitus and its association with mortality using machine learning in three independent cohorts from South Korea, Japan, and the UK: a model development and validation study. eClinicalMedicine. 2025;80:103069. doi:10.1016/j.eclinm.2025.103069.

[5] (Mahajan et al., 2022) Mahajan A, et al., Multi-ancestry genetic study of type 2 diabetes highlights the power of diverse populations for discovery and translation. Nat Genet. 2022;54(5):560-572. doi:10.1038/s41588-022-01058-3.

[6] (Ge et al., 2022) Ge T, Chen CY, Ni Y, Feng YA, Smoller JW. Development and validation of a trans-ancestry polygenic risk score for type 2 diabetes in diverse populations. Genome Med. 2022;14(1):70. doi:10.1186/s13073-022-01074-2.

[7] (Khera et al., 2018) Khera AV, et al., Genome-wide polygenic scores for common diseases identify individuals with risk equivalent to monogenic mutations. Nat Genet. 2018;50(9):1219-1224. doi:10.1038/s41588-018-0183-z.

[8] (Liu et al., 2021) Liu W, Zhuang Z, Wang W, Huang T, Liu Z. An Improved Genome-Wide Polygenic Score Model for Predicting the Risk of Type 2 Diabetes. Front Genet. 2021;12:632385. doi:10.3389/fgene.2021.632385.

[9] (Mars et al., 2020) Mars N, et al., Polygenic and clinical risk scores and their impact on age at onset and prediction of cardiometabolic diseases and common cancers. Nat Med. 2020;26:549-557. doi:10.1038/s41591-020-0800-0.

[10] (Kim et al., 2024) Kim NY, et al., The clinical relevance of a polygenic risk score for type 2 diabetes mellitus in the Korean population. Sci Rep. 2024;14:5749. doi:10.1038/s41598-024-55313-0.

[11] (Liu et al., 2025) Liu Y, Wang H, Gao T, Yan Y, Wang T, Zheng C, Zeng P. Influence and role of polygenic risk score in the development of 32 complex diseases. J Glob Health. 2025;15:04071. Available at: https://pmc.ncbi.nlm.nih.gov/articles/PMC11893022/.

[12] (Abdellaoui et al., 2023) Abdellaoui A, Yengo L, Verweij KJH, Visscher PM. 15 years of GWAS discovery: Realizing the promise. Am J Hum Genet. 2023;110(2):179-194. doi:10.1016/j.ajhg.2022.12.011.

[13] (Zicari et al., 2021) Zicari RV, et al., Z-Inspection: A Process to Assess Trustworthy AI. IEEE Transactions on Technology and Society. 2021;2(2):83-97. doi:10.1109/TTS.2021.3066209.

[14] (Koch et al., 2023) Koch S, Schmidtke J, Krawczak M, Caliebe A. Clinical utility of polygenic risk scores: a critical 2023 appraisal. J Community Genet. 2023;14(5):471-487. doi:10.1007/s12687-023-00645-z.

[15] (Kim and Lee, 2025) Kim NY, Lee S. XPRS: a tool for interpretable and explainable polygenic risk score. Bioinformatics. 2025;41(4):btaf143. doi:10.1093/bioinformatics/btaf143.

[16] (Lundberg and Lee, 2017) Lundberg SM, Lee S-I. A Unified Approach to Interpreting Model Predictions. Advances in Neural Information Processing Systems (NeurIPS). 2017.

[17] (Z-Inspection KOREA, 2025) Z-Inspection Initiative. KOREA Use-Case Main Document: Assessment of the XPRS Polygenic Risk Score Tool for Type 2 Diabetes. Unpublished internal report. 2025.

[18] (Steinthorsdottir et al., 2007) Steinthorsdottir V, Thorleifsson G, Reynisdottir I, et al., A variant in CDKAL1 influences insulin response and risk of type 2 diabetes. Nat Genet. 2007;39(6):770-775. doi:10.1038/ng2054.

[19] (Zicari et al., 2022) Zicari RV, et al., How to Assess Trustworthy AI in Practice. arXiv:2206.09887 [cs.CY]. 2022.

[20] (AI HLEG, 2019) High-Level Expert Group on Artificial Intelligence. Ethics Guidelines for Trustworthy AI. European Commission. 2019. Available at: https://ec.europa.eu/digital-single-market/en/news/ethics-guidelines-trustworthy-ai.

[21] (Robertson et al., 2019) Robertson LJ, Abbas R, Alici G, Munoz A, Michael K. Engineering-Based Design Methodology for Embedding Ethics in Autonomous Robots. Proc. IEEE. 2019;107(3):582-599. doi:10.1109/JPROC.2018.2889678.

[22] (NASA, 2021) Technology Readiness Level (TRL) as the foundation of Human Readiness Level (HRL). NASA Technical Reports. 2021. Available at: https://ntrs.nasa.gov/api/citations/20210012614/downloads/.

[23] (European Parliament, 2024) European Parliament and Council of the European Union. Regulation (EU) 2024/1689 laying down harmonised rules on artificial intelligence (AI Act). Official Journal of the European Union, L series, 12 July 2024.

[24] (NIA Korea, 2025) National Information Society Agency (NIA). Draft Enforcement Decree to the AI Basic Act. Republic of Korea. 2025. Available at: https://nia.or.kr/site/nia_kor/ex/bbs/View.do?cbIdx=99835&bcIdx=28600&parentSeq=28600.

[25] (National Assembly of Korea, 2025) National Assembly of the Republic of Korea. AI Basic Act (Framework Act on Artificial Intelligence). Enacted January 2025.

[26] (HUDERIA Methodology, 2024) Committee on Artificial Intelligence (CAI). Methodology for the Risk and Impact Assessment of Artificial Intelligence Systems from the Point of View of Human Rights, Democracy and the Rule of Law. Council of Europe, Strasbourg, 28 November 2024. Available at: https://rm.coe.int/cai-2024-16rev2-methodology-for-the-risk-and-impact-assessment-of-arti/1680b2a09f.

[27] (Council of Europe, 2024) Council of Europe Treaty Series - No. 225. Council of Europe Framework Convention on Artificial Intelligence and Human Rights, Democracy and the Rule of Law. Vilnius, 5.IX.2024. Available at: https://rm.coe.int/1680afae3c.

[28] (Zicari et al., 2024) Zicari RV, et al., Lessons Learned in Performing a Trustworthy AI and Fundamental Rights Assessment. arXiv:2404.14366. 2024. Available at: https://arxiv.org/pdf/2404.14366.

[29] (OHCHR, 2023) United Nations Office of the High Commissioner for Human Rights (OHCHR). The right to access to, to participate in, and to enjoy the benefits of scientific progress. Available at: https://www.ohchr.org/en/special-procedures/sr-cultural-rights/right-access-and-participate-science.

[30] (UN General Assembly, 1966) United Nations General Assembly. International Covenant on Civil and Political Rights (ICCPR). Adopted 16 December 1966, entered into force 23 March 1976. United Nations Treaty Series, vol. 999, p. 171.

[31] (Petersen, 2012) Petersen N. Human Dignity, International Protection. Max Planck Encyclopedia of Public International Law. Oxford University Press. Available at: https://opil.ouplaw.com/display/10.1093/law:epil/9780199231690/law-9780199231690-e809.

[32] (Ministry of Science and ICT, 2020) Ministry of Science and ICT. The National Guidelines for AI Ethics. Republic of Korea, 1 December 2020.

[33] (Schwarzerova et al., 2024) Schwarzerova J, Hurta M, Barton V, et al., A perspective on genetic and polygenic risk scores: advances and limitations and overview of associated tools. Brief Bioinform. 2024;25(3):bbae240. doi:10.1093/bib/bbae240.

[34] (Jayasinghe et al., 2024) Jayasinghe D, Eshetie S, Beckmann K, Benyamin B, Lee SH. Advancements and limitations in polygenic risk score methods for genomic prediction: a scoping review. Hum Genet. 2024;143(12):1401-1431. doi:10.1007/s00439-024-02716-8.

[35] (Wang et al., 2022) Wang Y, Tsuo K, Kanai M, Neale BM, Martin AR. Challenges and Opportunities for Developing More Generalizable Polygenic Risk Scores. Annu Rev Biomed Data Sci. 2022;5:293-320. doi:10.1146/annurev-biodatasci-111721-074830.

[36] (Campbell et al., 2021) Campbell JP, Mathenge C, Cherwek H, et al., Artificial Intelligence to Reduce Ocular Health Disparities. Transl Vis Sci Technol. 2021;10(3):19. doi:10.1167/tvst.10.3.19.

[37] (Elhussein et al., 2024) Elhussein A, Baymuradov U, NYGC ALS Consortium, et al., A framework for sharing of clinical and genetic data for precision medicine applications. Nat Med. 2024;30:3578-3589. doi:10.1038/s41591-024-03239-5.

[38] (Matsuyama et al., 2023) Matsuyama K, Kurihara C, Crawley FP, Kerpel-Fronius S. Utilization of genetic information for medicines development and equitable benefit sharing. Front Genet. 2023;14:1085864. doi:10.3389/fgene.2023.1085864.

[39] (Sorani et al., 2015) Sorani MD, Yue JK, Sharma S, Manley GT, Ferguson AR; TRACK TBI Investigators. Genetic data sharing and privacy. Neuroinformatics. 2015;13(1):1-6. doi:10.1007/s12021-014-9248-z.

[40] (MedlinePlus, 2023) MedlinePlus. What are the benefits and risks of direct-to-consumer genetic testing? MedlinePlus Genetics. 2023. Available at: https://medlineplus.gov/genetics/understanding/dtcgenetictesting/dtcrisksbenefits/.

[41] (Truong et al., 2023) Truong B, Hull LE, Ruan Y, et al., Integrative polygenic risk score improves the prediction accuracy of complex traits and diseases. medRxiv. 2023. doi:10.1101/2023.02.21.23286110.

[42] (Sun et al., 2021) Sun J, Wang Y, Folkersen L, et al., Translating polygenic risk scores for clinical use by estimating the confidence bounds of risk prediction. Nat Commun. 2021;12:5276. doi:10.1038/s41467-021-25014-7.

[43] (Misra et al., 2025) Misra A, Truong B, Urbut SM, et al., Instability of high polygenic risk classification and mitigation by integrative scoring. Nat Commun. 2025;16:1584. doi:10.1038/s41467-025-56945-0.

[44] (Rahnasto, 2023) Rahnasto J. Genetic data are not always personal, disaggregating the identifiability and sensitivity of genetic data. J Law Biosci. 2023;10(2):lsad029. doi:10.1093/jlb/lsad029.

[45] (Amorim et al., 2022) Amorim M, Silva S, Machado H, et al., Benefits and Risks of Sharing Genomic Data for Research. Int J Environ Res Public Health. 2022;19(14):8788. doi:10.3390/ijerph19148788.

[46] (Rehm et al., 2021) Rehm HL, Page AJH, Smith L, et al., GA4GH: International policies and standards for data sharing across genomic research and healthcare. Cell Genomics. 2021;1(2):100029. doi:10.1016/j.xgen.2021.100029.

[47] (Evangelou et al., 2018) Evangelou E, Warren HR, Mosen-Ansorena D, et al., Genetic analysis of over 1 million people identifies 535 new loci associated with blood pressure traits. Nat Genet. 2018;50:1412-1425. doi:10.1038/s41588-018-0205-x.

[48] (Kuchenbaecker et al., 2017) Kuchenbaecker KB, Hopper JL, Barnes DR, et al., Risks of Breast, Ovarian, and Contralateral Breast Cancer for BRCA1 and BRCA2 Mutation Carriers. JAMA. 2017;317(23):2402-2416. doi:10.1001/jama.2017.7112.

[49] (Harada-Shiba et al., 2023) Harada-Shiba M, Arai H, Ohmura H, et al., Guidelines for the Diagnosis and Treatment of Adult Familial Hypercholesterolemia 2022. J Atheroscler Thromb. 2023;30(5):558-586. doi:10.5551/jat.CR005.

[50] (Deignan et al., 2020) Deignan JL, Astbury C, Cutting GR, et al., CFTR variant testing: a technical standard of ACMG. Genet Med. 2020;22(8):1288-1295. doi:10.1038/s41436-020-0822-5.

[51] (Omrani et al., 2022) Omrani N, Rivieccio G, Fiore U, Schiavone F, Agreda SG. To trust or not to trust? An assessment of trust in AI-based systems. Technol Forecast Soc Change. 2022;181:121763.

[52] (Alhazmi, 2025) Alhazmi F. Understanding Physician Attitudes Toward AI in Clinical Decision-Making: Cross-Sectional Study. JMIR Form Res. 2025;9:e79730.

[53] (Tamori et al., 2022) Tamori H, Yamashina H, Mukai M, et al., Acceptance of the use of artificial intelligence in medicine among Japan's doctors and the public. JMIR Hum Factors. 2022;9(1):e24680.

[54] (Oh et al., 2019) Oh S, Kim JH, Choi SW, et al., Physician confidence in artificial intelligence: an online mobile survey. J Med Internet Res. 2019;21(3):e12422.

[55] (Shin, 2021) Shin D. The effects of explainability and causability on perception, trust, and acceptance: Implications for explainable AI. Int J Hum Comput Stud. 2021;146:102551.

[56] (Herzog et al., 2025) Herzog C, Blank S, Stahl BC. Towards trustworthy medical AI ecosystems. AI Soc. 2025;40(4):2119-2139.

[57] (Donohue et al., 2021) Donohue KE, Gooch C, Katz A, et al., Pitfalls and challenges in genetic test interpretation. Clin Genet. 2021;99(5):638-649. doi:10.1111/cge.13917.

[58] (Campbell et al., 2002) Campbell EG, Clarridge BR, Gokhale M, et al., Data Withholding in Academic Genetics. JAMA. 2002;287(4):473-480. doi:10.1001/jama.287.4.473.

[59] (Inouye et al., 2018) Inouye M, Abraham G, Nelson CP, et al., Genomic Risk Prediction of Coronary Artery Disease in 480,000 Adults. J Am Coll Cardiol. 2018;72(16):1883-1893. doi:10.1016/j.jacc.2018.07.079.

[60] (Antoniou et al., 2008) Antoniou AC, Spurdle AB, Sinilnikova OM, et al., Common breast cancer-predisposition alleles are associated with breast cancer risk in BRCA1 and BRCA2 mutation carriers. Am J Hum Genet. 2008;82(4):937-948. doi:10.1016/j.ajhg.2008.02.008.

[61] (Kim, 2020) Kim ES. Deep learning and principal-agent problems of algorithmic governance: The new materialism perspective. Technol Soc. 2020;63:101378.

[62] (ScienceDirect, 2024) ScienceDirect Topics. Genetic database, an overview. ScienceDirect. 2024. Available at: https://www.sciencedirect.com/topics/immunology-and-microbiology/genetic-database.

[63] (Fahed et al., 2020) Fahed AC, Wang M, Homburger JR, et al., Polygenic background modifies penetrance of monogenic variants for tier 1 genomic conditions. Nat Commun. 2020;11:3635. doi:10.1038/s41467-020-17374-3.

[64] (Fritzsche et al., 2023) Fritzsche MC, Akyuz K, Cano Abadia M, et al., Ethical layering in AI-driven polygenic risk scores. Front Genet. 2023;14:1098439. doi:10.3389/fgene.2023.1098439.

# APPENDIX A. SOCIO-TECHNICAL SCENARIOS TEMPLATE

1. **Aim of the system**
   Goal of the system, context, and why it is used.

2. **Actors** (*primary:* directly involved with the use of the AI system, and *secondary* and *tertiary* only indirectly involved with the use of the AI system)
   Who designed and implemented the system? Who has authorized the deployment of the system? Who is currently using the system? Who are the end users for this system? Who is directly influenced by decisions made by the system? Who is indirectly influenced by decisions made by the system? Who is responsible for this system?

3. **Actors' Expectation and Motivation**
   Why would the different groups of actors want the system? What are their expectations of the system behavior? What benefits are they expecting from using the system?

4. **Actors' Concerns and Worries**
   What problems/challenges can the actors foresee? Do they have concerns regarding the use of the system? What risks are they concerned about with the system? Are there any conflicts?

5. **Context where the AI system is used**

   What additional context information about the situation where the AI system is used? (e.g., urgency, budget constraints, for-profit, academic, conflicts, environmental). What are the potential future uses of the AI system?

6. **Interaction with the AI system**
   What is the intended interaction between the system and its users? If and how the 'human in control' aspect envisaged? Why is it like this?

7. **AI Technology used**
   Technical description of the AI system. An important part of considering AI trustworthiness is that it is robust, and if the technical description is not clear, this cannot be assessed.

8. **Clinical studies /Field tests**
   Was the system's performance validated in (clinical/field tests) studies? What were the results of these studies? Are the results openly available?

9. **Intellectual Property**
   What parts of the AI system are open access (if any)? What IP regulations need to be considered when assessing/disseminating the system? Does it contain confidential information that must not be published? What is and how to handle the IP of the AI and of the part of the system to be examined. Identify possible restrictions to the Inspection process, in this case, assess the consequences (if any). Define if and when Code Reviews are needed/possible.

10. **Legal framework**
    What is the legal framework for the use of the system? What special regulations apply? What are the data protection issues?" "Was the data aspect compliant with the GDPR?

11. **Ethics oversight and/or approval**
    Has the AI system already undergone an ethical assessment or other approval? If not - why not? If so, was this internal/external, volunteer/regulated, and what was covered? Did they get a waiver? Was there a clearing, but it was very light or internal and not considered sufficient?

# APPENDIX B LOG

...............................................................................

Workshop(s): date(s), Location(s): **GSDS, SNU, March 26, 2025**

For each workshop list of the names, affiliation and roles and expertise of each participants:

**Heejin Kim,** (Legal scholar)

**Seunggeun Lee,** (Biostatistics scholar)

**Yuna Park,** (Phd Student)

**Roberto V. Zicari** (Z-lead)

...................................................................................

Workshop(s): date(s), Location(s): **GSDS, SNU, April 9, 2025**

For each workshop list of the names, affiliation and roles and expertise of each participants:

**Heejin Kim,** (Legal scholar)

**Seunggeun Lee,** (Biostatistics scholar)

**Yuna Park,** (Phd Student)

**Roberto V. Zicari** (Z-lead)

**[Stephan Sonnenberg]** (Human Rights and Law)

...................................................................................

Workshop(s): date(s), Location(s): **GSDS, SNU, April 23, 2025**

For each workshop list of the names, affiliation and roles and expertise of each participants:

**Yuna Park,** (Phd Student)

**Roberto V. Zicari** (Z-lead)

[**Sira Maliphol** ( Technology and Society)]

...................................................................................

Workshop(s): date(s), Location(s): **Zoom april 29**

Kick off via Zoom of the project. Session recorded and link shared to the team

...................................................................................

Extra short Workshop(s): date(s), Location(s): **GSDS, SNU, April 30, 2025**

For each workshop list of the names, affiliation and roles and expertise of each participants:

**Heejin Kim,** (Legal scholar)

**Seunggeun Lee,** (Biostatistics scholar)

**Yuna Park,** (Phd Student)

**Roberto V. Zicari** (Z-lead)

..........................................................................................

Workshop(s): date(s), Location(s): **GSDS, SNU, May 7, 2025**

For each workshop list of the names, affiliation and roles and expertise of each participants:

**Seunggeun Lee,** (Biostatistics scholar)

**Yuna Park,** (Phd Student)

**Roberto V. Zicari** (Z-lead)

..........................................................................................

Workshop(s): date(s), Location(s):Zoom, May 13, 2025

For each workshop list of the names, affiliation and roles and expertise of each participants:

Seunggeun Lee, (Biostatistics scholar) presented the use case and answered a number of questions. The session was recorded and the link shared to the team.

..........................................................................................

May 17, 2025

**Vadim Pak**, Council of Europe, has joined our team.

..........................................................................................

June 4, 2025

In person meeting at SNU with **Heejin Kim, Yuna Park,** and **Roberto V. Zicari**

..........................................................................................

June 18, 2025

In person meeting at SNU between **Heejin Kim, Seunggeun Lee, Yuna Park, and Roberto V. Zicari**

..........................................................................................

*June 25, 2025*

General Zoom meeting

Participants:

**Roberto**

**Heejin**

**Elisabeth**

**Magnus**

**Pedro**

**Emilie**

**Yuna**

**Seunggeun**

**Stephan**

**Ralf**

**Sira**

……………………………………………………………………………………...

*June 26*

Zoom meeting of the WG1 Ethics/Law/Human Rights

Coordinator: **Elisabeth Hildt, Participants of the WG1**

**end Log**